\documentclass[preprint,12pt]{elsarticle}

\usepackage{amssymb}
\usepackage{amsmath, graphicx,amssymb,epsfig,psfrag,subfigure,bm,bbm,booktabs}
\usepackage{color}

\newcommand{\beq}{\begin{equation}}
\newcommand{\eeq}{\end{equation}}
\newcommand{\bpm}{\begin{pmatrix}}
\newcommand{\epm}{\end{pmatrix}}
\newcommand{\beqa}{\begin{eqnarray}}
\newcommand{\eeqa}{\end{eqnarray}}
\newcommand{\beqas}{\begin{eqnarray*}}
\newcommand{\eeqas}{\end{eqnarray*}}

\renewcommand{\d}{\mathrm{d}}

\newcommand{\pdhfrac}[2]{\mathchoice{\frac{#1}{#2}}{#1/#2}{#1/#2}{#1/#2}}

\newcommand{\pd}[2]{\pdhfrac{{\partial}#1}{{\partial}#2}}

\newcommand{\be}{\beta}

\def\XXint#1#2#3{{\setbox0=\hbox{$#1{#2#3}{\int}$ }
\vcenter{\hbox{$#2#3$ }}\kern-.6\wd0}}

\journal{Computer Methods in Applied Mechanics and Engineering}

\begin{document}

\begin{frontmatter}



\title{Generation of smoothly-varying infill configurations from a continuous menu of cell patterns and the asymptotic analysis of its mechanical behaviour}


\author[DUT1,PennState]{Dingchuan Xue}

\author[DUT1,DUT2,DUT3]{Yichao Zhu\corref{mycorrespondingauthor}}
\cortext[mycorrespondingauthor]{Corresponding authors}
\ead{yichaozhu@dlut.edu.cn}

\author[DUT1,DUT2,DUT3]{Xu Guo\corref{mycorrespondingauthor}}
\ead{guoxu@dlut.edu.cn}

\address[DUT1]{Department of Engineering Mechanics, Dalian University of Technology, Dalian, 116023, P. R. China}
\address[DUT2]{State Key Laboratory of Structural Analysis for Industrial Equipment, Dalian University of Technology}
\address[DUT3]{International Research Center for Computational Mechanics, Dalian University of Technology}
\address[PennState]{Department of Engineering Science and Mechanics, Pennsylvania State University,
University Park, Pennsylvania 16802, USA}

\begin{abstract}
  We here introduce a novel scheme for generating smoothly-varying infill graded microstructural (IGM) configurations from a given menu of generating cells. The scheme was originally proposed for essentially improving the variety of describable configurations in a modified asymptotic homogenisation-based topology optimisation framework \cite{ZhuYC_JMPS2019} for fast IGM design. But the proposed scheme, after modification, also demonstrates its unique values in two aspects of applications. First, it provides a fairly simple way of generating an IGM configuration continuously patching any given cell configurations. Second, it tenders a straightforward mean for decorating microstructures on a given manifold. We will further show that the form of topology description function given here effectively offers a platform for unifying most existing approaches for IGM generation. Fuelled by asymptotic analysis of the mechanical behaviour of the resulting IGM configurations, a topology optimisation scheme for compliance minimisation is introduced. We will finally show that, the use of the present scheme helps reduce the compliance value of an optimised structure by nearly a half, if compared with that from the original framework \cite{ZhuYC_JMPS2019}.
\end{abstract}

\begin{keyword}


Infill lattice structures \sep Graded microstructures \sep Topology optimisation \sep Asymptotic analysis \sep Topology description function
\end{keyword}

\end{frontmatter}


\section{Introduction}
Infill graded microstructural (IGM) configurations have shown their vast engineering potentials in achieving exceptional material/structural properties \cite{Lakes_Nature1993,DongHW_JMPS2017,Sigmund_APL1996,Kushwaha_PRL1993,LiuC_PRAppl2015, Aage_Nature2017}. For IGM design, early treatments encompass the use of asymptotic homogenisation (AH) approaches \citep{Bensoussan1978}. The basic idea is to treat a spatially periodic configuration as a continuum, whose elasticity tensor at every point is calculated through solving a single set of partial differential equations (PDEs) defined over one constituting cell. The homogenised formulation was then adopted for spatially-varying infill configurations, where every macroscopic pixel is associated with an individual cell \citep{Bendsoe_CMAME1988,Bendsoe_StructOpt1989,Sigmund_IJSS1994}. The method was further extended to other cases, such as concurrent multi-scale design \cite{Rodrigues_SMO2002,LiuL_CompStruct2008}, multi-functional design \citep{NiuB_SMO2009,DengJD_SMO2013,YanJ_CompMech2016} and problems concerning load uncertainties \cite{DengJD_SMO2017}. However, several limiting issues arise against the treatments based on conventional AH formulation. For instance, the smoothness across neighbouring cells may not be guaranteed. Moreover, the asymptotic analytical results may become inaccurate, since the presumption of microstructural periodicity no longer holds.

For improvement, various types of methods for IGM representation and stress analysis have been proposed. An intuitive way of thinking is to pile up building-block structures, whose topology description functions (TDFs) originate from a single set of continuous functions. This can be achieved either explicitly or implicitly. The explicit way involves cell configurations composed of parameterised members \citep{ZhouSW_JMaterSci2008,Radman_JMaterSci2013,Radman_CompMaterSci2014,ChengL_CMAME2019, LiQH_CompStruct2019,WangC_SMO2020}. Thus an IGM configuration is generated by letting the corresponding parameters vary gradually (but discretely) in space. The implicit treatment is usually achieved through the introduction of a level-set function, whose multiple (but discrete) level-set contours are used to characterise the boundaries of a family of prototype cells \cite{WangYQ_CMAME2017,Gao_AES2018,ZhangY_CMS2018,ZhangY_SMO2019}. By either mean, graded change in microstructure should be achieved, but several challenging issues still persist. First, slight non-smoothness across the boundaries of neighbouring cells is still an unresolved issue, since only discrete values of the controlling functions are concerned. Second, the variation of the microstructural infill patterns is only permitted to take place along directions parallel to cell edges. This greatly squeezes the resulting IGM variety.

Recently, a number of novel approaches underpinning fast IGM design are proposed, where smoothness connections within IGM configurations can be fully guaranteed. Interestingly, the restriction on oriented rectangle/cuboid-shape cells is also removed in these recently proposed approaches. Besides simulations directly conducted on fine scales
\cite{Alexandersen_CMAME2015,LiuC_JAM2017}, the novelly introduced treatments can be roughly classified into three categories: 1) conformal-mapping-based approaches; 2) the de-homogenisation approaches; 3) the modified asymptotic homogenisation topology optimisation (AHTO plus) approaches.

The conformal-mapping-based approaches \cite{Vogiatzis_CMAME2018} mainly concern decorating microstructural configurations on a given surface in three-dimensional space. In this scenario, conforming mapping is used to ensure that the local orthogonality in microstructure is preserved as a planar microstructure is deployed on the prescribed surface. As a cost, the Ricci flow equation, a useful differential equation defined on manifolds, needs to be numerically solved to a steady state.

The de-homogenisation methods \cite{Groen_IJNME2018,Groen_CMAME2019,Allaire_CompMathAppl2019, WuJ_IEEE2019, GeoffroyDonders_JCP2020} focus on cells of laminate configurations, which theoretically guarantee the solution optimality for most cases in compliance optimisation. They first determine an optimised distribution of the laminate volume fraction and orientation (on a coarser grid). Then the actual microstructure is resolved by solving another optimisation problem projecting the laminate configurations in alignment with the desired directions of the local principal stresses \cite{Pantz_SIAMJCO2008}.

The AHTO plus approaches \cite{ZhuYC_JMPS2019} describe an IGM configuration with function composition. The generated IGM configuration can always be mapped to a spatially periodic configuration (in a fictitious space) by means of a continuous mapping function. Then the TDF of the IGM configuration is obtained through the function composition of the mapping function with the TDF describing the cell configuration (of one period in the fictitious space). Under this setting, asymptotic analysis is carried out for deriving a homogenised formulation governing the mechanical behaviour of the resulting IGM configuration. Compared with the above-mentioned two methods, no post-processing procedure, such as solving the Ricci flow equation or conducting projecting operations, is needed in the AHTO plus framework. Nevertheless, other challenging issues arise over the following three aspects. Firstly, since the constituting cells (of parallelogram/parallelepiped shape) originate from a single generating cell, thus they must be self-similar to each other. This dramatically reduces the IGM variety in concern. For instance, spatial change in microstructural volume fraction is still disabled. Secondly, the number of design variables, which parameterise the controlling function, is relatively small under the AHTO plus framework. Thus the identification of an appropriate set of design variables becomes a critical issue, which has yet been properly addressed. Finally, the derived homogenisation formulation involves solving various PDE systems at individual macroscopic points, generating a huge computational burden. Hence finding an efficient way of solving these PDEs \cite{XueDC_arxiv2019} is also a key factor affecting the applicability of the method for IGM design.

The present article is aimed for addressing the first aforementioned issue, that is, to essentially improve the IGM variety under the AHTO plus framework. Compared to the original treatment in \cite{ZhuYC_JMPS2019} where local microstructures have to stem from a single generating cell, here we consider a continuous menu of generating cell configurations defined in a fictitious space, which can be systematically identified by the continuous level sets of a single function, say, $\phi(\cdot)$, termed as a ``structural generating function'' in this article. Then a macroscopic continuous function defined in the actual space, say $\zeta(\mathbf{x})$, is introduced to indicate which level set it takes around point $\mathbf{x}$. Here the function $\zeta(\mathbf{x})$ is termed as a ``level-set indicator'' function. Incorporating $\zeta(\mathbf{x})$ with the original function composition treatment, a TDF associated with an IGM configuration is thus introduced, and the ``graded microstructural'' behaviour is carried out by the combinative effects of the following two sets of operations: manipulation of a single cell and transition among cells. The action of cell manipulation involves 1) cell rescaling; 2) cell rotation; 3) cell distortion in an iso-volumetric manner. Mathematically, these operations over a single cell can be carried out by manipulating the local Jacobian matrix, which consists of the spatial derivatives of the mapping function. Such an idea for IGM representation originates from the use of level-set contours for modelling dislocation loops in crystalline materials \cite{XiangY_JMPS2009,ZhuYC_JMPS2015,Chapman_SIAP2016}. It bears certain similarity with the treatment by \cite{WangYQ_CMAME2017}, but essential differences exist. For instance, level sets of discrete values are used in \cite{WangYQ_CMAME2017}, but continuous level sets are considered here. Moreover, local rotation and distortion of the generating cells are not taken into account by \cite{WangYQ_CMAME2017}, resulting in pixel-wise microstructures only. This is, however, no longer a limiting issue if the present scheme is implemented. In fact, we will show that the TDF form given by Eq.~\eqref{TDF_general} effectively offers a platform for unifying most existing treatments for IGM generation, and detailed discussion is carried out in Sec.~\ref{Sec_unification}. The present algorithm also sees its generalised applications in two important aspects. Firstly, it offers a simple treatment for continuously patching arbitrarily given cell configurations, which are termed as ``seminal cell'' configurations in the present article. Secondly, it breeds a general algorithm for generating an IGM surface.

The IGM configurations obtained here inherit all the attractive features from that generated based on the original AHTO plus framework, such as microstructural smoothness and simple forms in TDFs. Moreover, the corresponding asymptotic formulation for stress analysis is also available, and its accuracy is further shown against fine-scale simulation results. Under the present framework, one can monitor an IGM configuration by varying the mapping function $\mathbf{y}(\mathbf{x})$, the level-set indicator function $\zeta(\mathbf{x})$ and the structural generating function $\phi(\cdot)$. Mathematically, this can be fulfilled by a topology optimisation formulation with the associated sensitivity analysis results given. All these should pave way for adapting the AHTO plus framework for fast digital design of IGM configurations.

The rest of the article is arranged as follows. In Sec.~\ref{Sec_AHTO_review}, the original AHTO plus method is briefly revisited with the challenging issues identified. In Sec.~\ref{Sec_spatial_variable_volume}, our algorithm for IGM generation from a continuous menu of cell configurations is introduced, and two of its generalised applications are discussed: 1) IGM generation continuously patching different seminal cells; 2) IGM generation on a given surface. The section concludes with discussion over the unification of the present methods for IGM representation under the present framework. In Sec.~\ref{Sec_homogenisation}, the asymptotic formulation for analysing IGM mechanical behaviour is firstly introduced. It is followed by the corresponding topology optimisation formulation, along with the associated sensitivity analysis results. In Sec.~\ref{Sec_numerical_examples}, two-dimensional examples will be firstly presented with several key issues of the algorithm addressed. Three-dimensional cases will also be shown in brief in the end of Sec.~\ref{Sec_numerical_examples}. The article concludes with further discussion in Sec.~\ref{Sec_conclusion}.

\section{IGM description under the original AHTO plus framework and challenging issues\label{Sec_AHTO_review}}
\subsection{IGM description under the original AHTO plus framework}
A graded microstructural configuration filling a certain domain, as shown in the bottom-left panel of Fig.~\ref{GMs}, normally involves at least two length scales:
\begin{figure}[!ht]
  \centering
  \includegraphics[width=.75\textwidth]{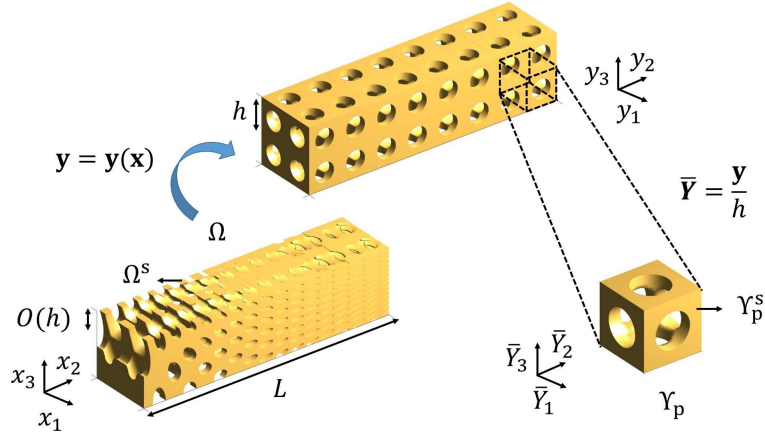}
  \caption{An IGM configuration generated through function composition by Eq.~\eqref{Map} under the original AHTO plus framework.}
  \label{GMs}
\end{figure}
a microscopic length scale on which the structural details are fully resolved, and a macroscopic length scale on which the trend in microstructural variation is observed. Here we characterise the ``microscale'' by a length parameter $h$, and the ``macroscale'' by the domain size $L$. Practically, we have
\beq\label{scale_separation}
\epsilon = \frac{h}{L} \ll 1.
\eeq

The mathematical representation of a structural configuration is usually realised through the introduction of a topology description function, denoted by $\chi(\mathbf{x})$ here. For a porous region $\Omega$ with the solid parts occupying a domain of $\Omega^{\text{s}}$, its associated TDF can be defined by
\begin{equation} \label{TDF_def}
\chi(\mathbf{x}) \left\{\begin{aligned}
& \ge 0,\quad &&\mathbf{x}\in\Omega^\text{s};\\
& <0, &&\mathbf{x}\in\Omega\setminus\Omega^\text{s}.
\end{aligned}\right.
\end{equation}
As for an IGM configuration, evaluating its associated TDF piece by piece on the fine scale is sometimes too time-consuming. For improvement, a treatment using function composition has been introduced \cite{Pantz_SIAMJCO2008,LiuC_JAM2017}, and further developped \cite{ZhuYC_JMPS2019}. The key is to make use of the smoothly-varying property of an IGM configuration.

As shown in Fig.~\ref{GMs}, a macroscopically smooth mapping function $\mathbf{y}=\mathbf{y}(\mathbf{x})$ is firstly introduced, so as to map an IGM configuration into a spatially \textit{periodic} one in a fictitious space, whose constituting cells are uniform and of a same length scale $h$. Then one may pick any cell (of rectangle/cuboid shape in the fictitious space) as a generating cell, and non-dimensionalise it into a region of unit dimensions, say $\Upsilon_{\text{p}} = \left[-1/2,1/2\right]^n$ with $n$ denoting the dimensionality number (as shown in the bottom-right panel of Fig.~\ref{GMs}). Now a TDF $\chi^\text{p}(\bar{\mathbf{Y}})$, $\bar{\mathbf{Y}}\in\left[-1/2,1/2\right]^n$ can be assigned to represent the structural configuration within the (non-dimensional) fictitious unit cell. Finally, the TDF $\chi(\mathbf{x})$ characterising the actual porous region $\Omega$ should be obtained by a composition of the (macroscopically smooth) mapping function $\mathbf{y}(\mathbf{x})$ and the TDF associated with the (fictitious) porous region in $\Upsilon_{\text{p}}$:
\beq\label{Map}
\chi(\mathbf{x})=\chi^\text{p}\left(\frac{\mathbf{y}(\mathbf{x})}{h}\right).
\eeq

The TDF generated through function composition based on Eq.~\eqref{Map} has shown many attractive features. Firstly, $\chi(\mathbf{x})$ by Eq.~\eqref{Map} is a continous function, because $\mathbf{y}(\mathbf{x})$ is smooth and $\chi^{\text{p}}(\cdot)$ is periodic over its input. Thus the resulting IGM configuration always takes a smoothly-varying profile. Secondly, Eq.~\eqref{Map} has a simple form, and this facilitates its implementation in a computer aided design (CAD) platform. Thirdly, it overcomes the limitation that the actual building blocks have to be of rectangle/cuboid shape. Actually, a variety of IMG configurations can be described by using Eq.~\eqref{Map}, as demonstrated in \cite{ZhuYC_JMPS2019}. Fourthly, Eq.~\eqref{Map} poses low demand on digital resources needed for memorising it. In theory, the mapping function $\mathbf{y}(\mathbf{x})$ can be defined on a coarser grid, and the TDF $\chi^{\text{p}}(\cdot)$ which is of high resolution spans only a single cell. Finally, a homogenisation formulation used for analysing the mechanical behaviour of an IGM configuration generated by Eq.~\eqref{Map} can be derived through mathematically rigorous asymptotic analysis \cite{ZhuYC_JMPS2019}, as $\epsilon$ tends to 0. Thus the stress analysis of an IGM configuration can still be performed at a coarse-grained level, and this renders significantly bring down the requirement on computational resources, when compared to simulations directly conducted on a fine scale.

\subsection{Challenging issues}
Albeit many attractive properties as mentioned above, the TDF generated by function composition through Eq.~\eqref{Map} still sees severe limitation. For instance, it struggles to describe an IGM configuration, where the microstructural volume fraction varies in space. This is because the mapping function $\mathbf{y}(\mathbf{x})$ simply rescales, rotates and distorts a single generating cell configuration. Thus the volume fraction stays unchanged after being transformed by $\mathbf{y}(\mathbf{x})$. For a detailed illustration over this point, one may refer to Appendix B in \cite{ZhuYC_JMPS2019}. Thus the present study is motivated: how to expand the coverage of the AHTO plus framework for describing more general IGM configurations.

\section{Representation of general IGM configurations\label{Sec_spatial_variable_volume}}
In this section, we will first discuss on how to extend the function composition treatment described in Eq.~\eqref{Map} for a much more general use. Then two of its generalised applications will be discussed in depth. We first introduce an algorithm for generating IGM configurations smoothly patching various seminal cells. Then the formulation is further extended for representing graded microstructures decorated on a given manifold. Finally, we will show that the present formulation effectively offers a theoretical framework that unifies various ways of IGM representation in literature.

\subsection{General formulation\label{Sec_representation}}
For the purpose of greatly improve the variety of describable IGM configurations using the function composition treatment, the TDF given by Eq.~\eqref{Map} is now modified as
\begin{equation}\label{TDF_general}
\chi(\mathbf{x}) = \phi\left(\frac{\mathbf{y}(\mathbf{x})}{h}\right) - \zeta(\mathbf{x}),
\end{equation}
where $\phi(\cdot)$, termed as the structural generating function in this article, is periodic over its input; $\zeta(\mathbf{x})$ is another smooth function in $\mathbf{x}$. A similar treatment has also been used in \cite{GuoX_CMAME2013} to account for possible boundary perturbations in topology optimization considering shape uncertainties.

\begin{figure}[!ht]
  \centering
  \includegraphics[width=.75\textwidth]{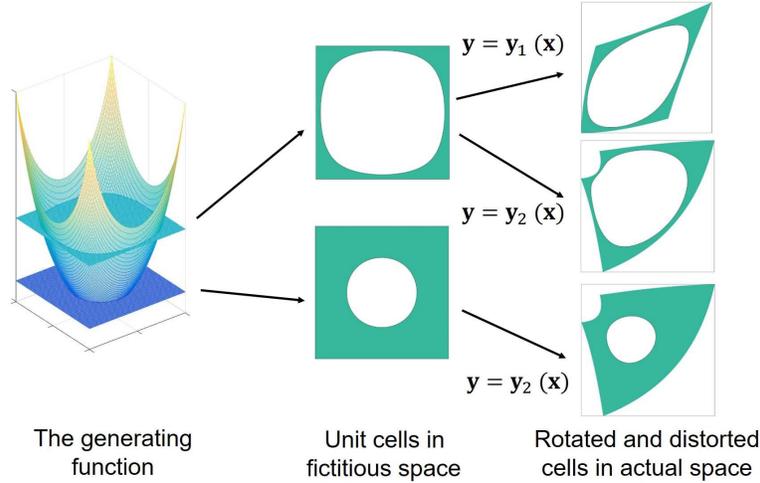}
  \caption{A continuous menu of generating cells are stored in a structural generating function by all its level-set contours. In the actual space, one should indicate which level set is of interest identified by $\zeta(\mathbf{x})$, and how the cell is rescaled, rotated, and iso-volumetrically distorted by $\mathbf{y}(\mathbf{x})$.}
  \label{Fig_illustration}
\end{figure}

The TDF defined by Eq.~\eqref{TDF_general} can be rationalised in a viewpoint of level-set functions. In a traditional level-set-based topology optimisation framework \cite{WangYu_CMAME2003,Allaire_JCP2004}, a level-set function is defined within the whole design domain, where only its zero-level-set contour has a geometric meaning of representing the structural boundaries. This idea is now generalised through Eq.~\eqref{TDF_general}, where all level-set values are geometrically meaningful. Practically, one may introduce a function $\phi(\bar{\mathbf{Y}})$, which is periodic over the fictitious non-dimensional domain of unit dimension measured by $\bar{\mathbf{Y}}$. As shown in Fig.~\ref{Fig_illustration}, its corresponding level-set contour of height $\zeta$, given by $\phi(\bar{\mathbf{Y}}) = \zeta$, specifies a certain structural boundary. Since $\phi(\bar{\mathbf{Y}})$ is continuous, the change in boundary profiles as $\zeta$ varies should be smooth, too. Hence a continuous menu of generating cell patterns are obtained in the fictitious (non-dimensional) space measured by $\bar{\mathbf{Y}}$. Then for a point $\mathbf{x}$ in the actual space, one needs to specify $\chi^{\text{p}}(\cdot)$ and $\zeta(\cdot)$. Here we let $\chi^{\text{p}}\left(\frac{\mathbf{y}}{h}\right) = \phi\left(\frac{\mathbf{y}}{h}\right)$. Thus Eq.~\eqref{TDF_general} indicates that the actual microstructural configuration around $\mathbf{x}$ is generated by manipulating the generating cell identified by the level-set contour of $\phi(\cdot)$ of height $\zeta=\zeta(\mathbf{x})$.

The IGM configurations generated from Eq.~\eqref{TDF_general} naturally inherit all the attractive features from the original treatment with function composition by Eq.\eqref{Map}. For instance, the resulting IGM configuration varies smoothly in space, because $\mathbf{y}(\mathbf{\mathbf{x}})$ and $\zeta(\mathbf{x})$ are both continuous over their inputs and $\chi^{\text{p}}(\cdot)$ is periodic over its input.

The transition between generating cells is thus enabled, and so does the volume change in microstructures. To formulate the microstructural volume fraction, we first identify the correlation between the contour of $\phi(\bar{\mathbf{Y}})$ of height $\zeta$ and the volume fraction associated with the corresponding generating cells. Mathematically, this is given by
\begin{equation} \label{vol_cell}
g(\zeta) = \int_{\Upsilon_{\text{p}}} H(\phi(\bar{\mathbf{Y}}) - \zeta) \, \d \bar{\mathbf{Y}},
\end{equation}
where $H(\cdot)$ is the Heaviside function. It is noted that the correlation formulated by Eq.~\eqref{vol_cell} is already set as the continuous menu of generating cells are present, which proceeds the representation of the final IGM configurations. Besides, cell manipulation, involving rescaling, rotation and iso-volumetric distortion, carried out through $\mathbf{y}(\mathbf{x})$, does not alter the corresponding local volume fraction. Thus the microstructural volume fraction around a spatial point $\mathbf{x}$ solely depends on the value of the level-set indicator function $\zeta(\mathbf{x})$, but is independent of $\mathbf{y}(\mathbf{x})$. Therefore, the overall volume fraction of an IGM configuration characterised by Eq.~\eqref{TDF_general} is formulated by
\begin{equation} \label{vol_fraction}
\frac{|\Omega^{\text{s}}|}{|\Omega|} = \int_{\Omega} g(\zeta(\mathbf{x}))\, \d \mathbf x,
\end{equation}
where $g(\cdot)$ is given by Eq.~\eqref{vol_cell}; ``$|\cdot|$'' denotes the volume occupied by the corresponding domain.

In theory, the significantly expanded IGM describability by Eq.~\eqref{TDF_general} is demonstrated. In the following, we will further explore its applicability in a more practically meaningful context.

\subsection{IGM generation from seminal cell patterns}
\begin{figure}[!ht]
\centering
\subfigure[A smiling face]{\includegraphics[width=.22\textwidth]{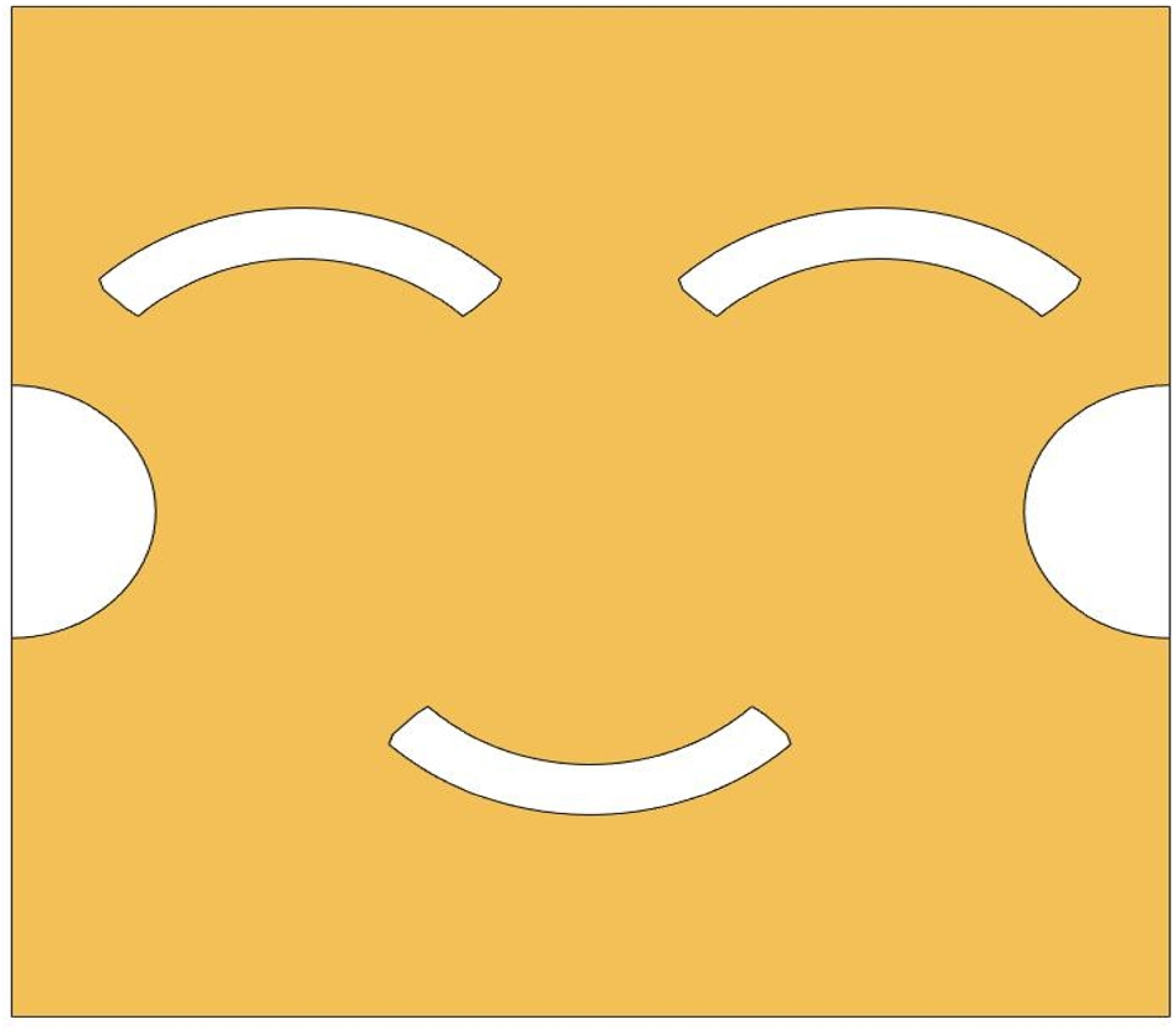}}
\qquad\qquad\qquad
\subfigure[A weeping face]{\includegraphics[width=.22\textwidth]{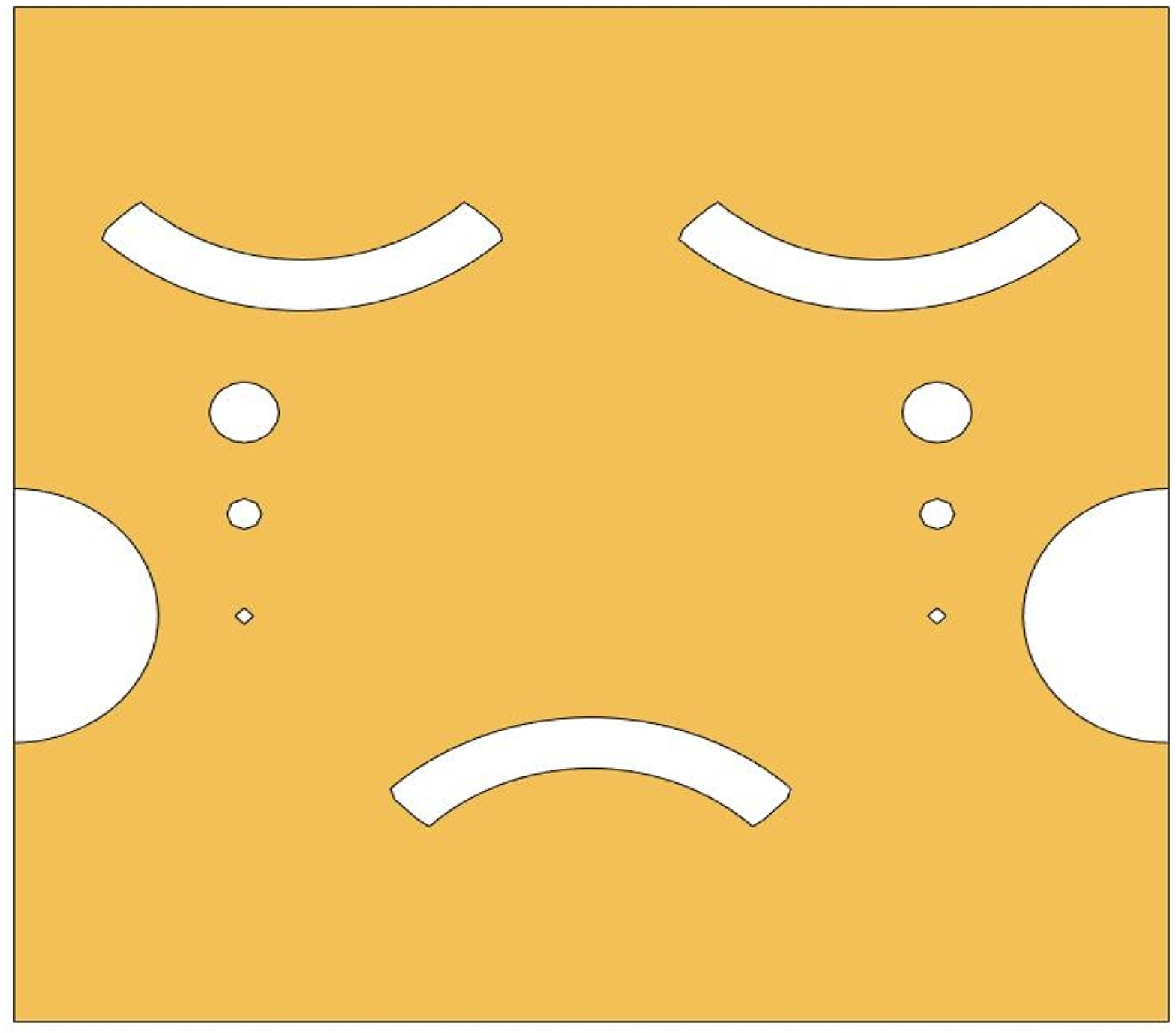}}
\caption{Two seminal cells: (a) a smiling face; (b) a weeping face.}
\label{expression_prototype}
\end{figure}

A number of microstructural configurations have been proposed for achieving exceptional functions \cite[e.g.][]{Sigmund_APL1996,Frenzel_Science2017}. One extended application of Eq.~\eqref{TDF_general} is to smoothly transit between cells outputting different high-end performances. For instance, two seminal cell structures are present: a smiling face and a weeping face, as shown in Fig.~\ref{expression_prototype}.
Then the question is how to design an IGM configuration where the facial expression continuously turns from smiling to weeping. Note that this is by no means a trivial task. For instance, the two seminal structures are topologically different. Hence it is of difficulty to use a single set of parameters to consistently express both configurations. Besides, the heights of ``cheeks'' of the two faces are different. Using spatially discrete parameters to patch them, e.g. \cite{WangYQ_CMAME2017,ChengL_CMAME2019}, may result in persistent mismatch between neighbouring cells.

Suppose the two facial configurations shown in Fig.~\ref{expression_prototype} are identified by the zero-level-set contours of two functions, $\phi_1(\cdot)$ and $\phi_2(\cdot)$, respectively. Then smooth connection between the two configurations should be obtained through interpolation between them. Hence we first construct a structural generating function by
\begin{equation} \label{phi_interpolation}
\phi(\bar{\mathbf{Y}};\zeta) = (1-\zeta) \phi_1(\bar{\mathbf{Y}}) + \zeta (\phi_2(\bar{\mathbf{Y}}) + 1),
\end{equation}
where $\zeta$ is the interpolation parameter. Compared to the structural generating function constructed in the previous subsection, $\phi$ defined by Eq.~\eqref{phi_interpolation} contains an extra parameter $\zeta$. This time, the structural boundaries of the generating cell patterns are still identified by the implicit relationship given by $\phi(\bar{\mathbf{Y}};\zeta) = \zeta$. It is checked that when $\zeta=0$, $\phi(\bar{\mathbf{Y}};\zeta) = \zeta$ becomes $\phi_1(\bar{\mathbf{Y}}) = 0$, corresponding to the structural boundary of the seminal cell given by Fig.~\ref{expression_prototype}(a). So on for the case when $\zeta=1$, corresponding to the weeping facial configuration identified by $\phi_2(\bar{\mathbf{Y}})=0$.

By first letting $\zeta=\zeta(\mathbf{x})$ in Eq.~\eqref{phi_interpolation} and incorporating it into Eq.~\eqref{TDF_general}, we derive the overall TDF associated with an IGM configuration by
\begin{equation} \label{chi_interpolation}
\chi(\mathbf{x}) = (1-\zeta(\mathbf{x})) \phi_1\left(\frac{\mathbf{y}(\mathbf{x})}{h}\right) + \zeta(\mathbf{x}) \phi_2\left(\frac{\mathbf{y}(\mathbf{x})}{h}\right).
\end{equation}
Although coming from Eq.~\eqref{TDF_general}, the TDF expression defined by Eq.~\eqref{chi_interpolation} has a clear mathematical meaning. It is generated through direct interpolation between the TDFs identifying the two seminal cell configurations, which is controlled by the level-set indicator function $\zeta(\mathbf{x})$.

\begin{figure}[!ht]
  \centering
  \includegraphics[width=.8\textwidth]{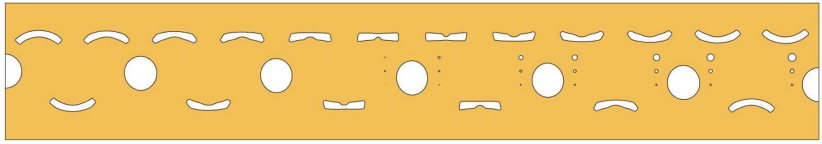}
  \caption{A smoothly-varying expression from smile to weeping.}
  \label{Face}
\end{figure}

With reference to Eq.~\eqref{chi_interpolation}, Fig.~\ref{Face} is generated, where the mapping function is set $\mathbf{y}=\mathbf{x}$ with $h=1$, and $\zeta(\mathbf{x})$ is a linear function of $x_1$, the spatial variable indicating the horizontal direction.
Note that tiny structural evolution takes place within every single cell. Thus the cells constituting the final IGM configuration are normally not exactly the same as any of the generating cells. This can be visualised in the third face to the left in Fig.~\ref{Face}, where tear drops only appear beneath its left eye. This again addresses the key different feature of the present method, as compared with that in \cite{WangYQ_CMAME2017}, where the constituting cells have to coincide with one of the prototype cells and tiny mismatch persists across cell boundaries.

\begin{figure}[!ht]
  \centering
  \includegraphics[width=.75\textwidth]{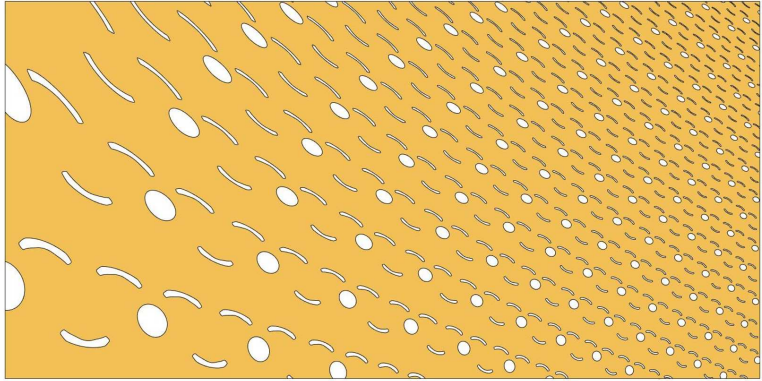}
  \caption{A wall of IGM configuration with mixed expressions.}
  \label{Face_2D}
\end{figure}

When the effect from the mapping function $\mathbf{y}(\mathbf{x})$ is taken into account, a wall of IGM configuration with mixed expressions, for instance, can be generated, as shown in Fig.~\ref{Face_2D}.
As a mapping treatment is now involved, several faces get rotated or distorted in the output IGM configuration.

The treatment can be extended to smoothly connect a number of seminal configurations. Suppose we have $M$ seminal cell configurations, represented by the zero level sets of $\phi_1$, $\cdots$, $\phi_i$, $\cdots$, $\phi_M$, respectively. Now we introduce a number set $\{\zeta_{i=1}^M\}$, and our goal is to generate a TDF such that it becomes $\phi_i(\cdot)$ when the interpolation parameter $\zeta$ equals $\zeta_i$. This can be achieved in a similar sense as Lagrangian interpolation, and a general TDF expression is thus defined by
\begin{equation} \label{chi_interpolation_s}
\chi(\mathbf{x}) = \sum_{i=1}^M \alpha_i \cdot \phi_i\left(\frac{\mathbf{y}(\mathbf{x})}{h}\right),
\end{equation}
where the coefficients $\alpha_i$ are formulated by
\begin{equation} \label{coefficient_alpha}
\alpha_i = \frac{(\zeta(\mathbf{x})-\zeta_1)\cdots(\zeta(\mathbf{x})-\zeta_{i-1}) (\zeta(\mathbf{x})-\zeta_{i+1})\cdots (\zeta(\mathbf{x})-\zeta_M)}{(\zeta_i-\zeta_1) \cdots(\zeta_i-\zeta_{i-1})(\zeta_i-\zeta_{i+1})\cdots(\zeta_i-\zeta_M)}.
\end{equation}

It can be checked that when $\zeta(\mathbf{x}) = \zeta_i$, the solid region is identified by $\phi_i(\cdot)>0$, which actually corresponds to the $i$-th seminal cell configuration. Therefore, the proposed method can be used for generating IGM configurations from any given set of seminal cell patterns.

In fact, Eq.~\eqref{chi_interpolation_s} may be extended to a more general form for describing a continuous menu of parameterised cell patterns, say identified by the generating function $\phi(\bar{\mathbf{Y}};\beta)$, where $\beta\in[\underline{\beta},\bar{\beta}]$ is a parameter. The corresponding TDF can be given by
\begin{equation} \label{chi_convolution}
\chi(\mathbf{x}) = \int_{\underline{\beta}}^{\bar{\beta}} \alpha(\zeta(\mathbf{x});\beta) \phi\left(\frac{\mathbf{y}(\mathbf{x})}{h};\beta\right)\,\d \beta.
\end{equation}
By this mean, the pool of generating cells can be enriched by ``mixing'' seminal cells at a certain ratio given by $\alpha(\zeta(\mathbf{x});\beta)$.

\subsection{Generation of surfaces with smoothly-varying microstructures}
The method can be naturally generalised for deploying microstructures on a given manifold. A surface in three dimensions can be parameterised by two free variables, say, $u$ and $v$, and it can thus be represented by
\begin{equation} \label{surface_x}
\mathbf{x}^{\text{m}} = \mathbf{f}^{\text{m}}(u,v),
\end{equation}
where $\mathbf{x}^{\text{m}}$ are the vector containing the coordinates of the points lying on the manifold; $\mathbf{f}^{\text{m}}$ is a vector-valued function from $\mathbb{R}^2$ to $\mathbb{R}^3$. Here a superscript ``$^{\text{m}}$'' is attached to a variable indicating its affiliation with a manifold. Now our target is to decorate it with smoothly-varying microstructures. For implementing the proposed method, we need to determine the TDFs of the generating cells and the mapping function $\mathbf{y}(\mathbf{x})$.

This time, the (continuous) menu of generating cells should be planar structures in two-dimensional space, and suppose they are given by a periodic structural generating function $\zeta=\phi(y_1,y_2)$, where the level-set of value $\zeta\ge0$ identifies a (two-dimensional) cell configuration. Then we consider how to construct the mapping function so as to decorate the menu of cells on the manifold given by Eq.~\eqref{surface_x}. It takes the following steps.

First, since $\mathbf{y}(\mathbf{x})$ is actually a map of $\mathbb{R}^3\rightarrow\mathbb{R}^3$, an extra degree of freedom (in conjunction with $u$ and $v$), say $w$, should be introduced. A natural way of thinking is to let it be in alignment with the surface normal $\mathbf{n}$, which is mathematically given by
\begin{equation} \label{surface_normal}
\mathbf{n} = \frac{\mathbf{f}^{\text{m}}_{,1} \times \mathbf{f}^{\text{m}}_{,2}}{|\mathbf{f}^{\text{m}}_{,1} \times \mathbf{f}^{\text{m}}_{,2}|},
\end{equation}
where $\mathbf{f}^{\text{m}}_{,i}$, $i=1$ and $2$, denote the derivative of each entry of $\mathbf{f}^{\text{m}}$ with respect to its $i$-th entry (e.g. $\mathbf{f}^{\text{m}}_{,1} = \frac{\partial \mathbf{f}^{\text{m}}}{\partial u}$), and the symbol ``$\times$'' denotes the cross product of two vectors. Then a map between $(u,v,w)$ and all points around the manifold is established:
\begin{equation} \label{map_surface0}
\mathbf{x} = \mathbf{x}^{\text{m}}(u,v) + \left.\frac{\mathbf{f}^{\text{m}}_{,1} \times \mathbf{f}^{\text{m}}_{,2}}{|\mathbf{f}^{\text{m}}_{,1} \times \mathbf{f}^{\text{m}}_{,2}|}\right|_{(u,v)} w,
\end{equation}
where the subscript $(u,v)$ indicates the two inputs of the corresponding function. Then we choose $y_1=u$, $y_2=v$ and $y_3=w$. A mapping function from the fictitious space of $\mathbf{y}$ to the actual space of $\mathbf{x}$ is defined:
\begin{equation} \label{map_surface}
\mathbf{x} \triangleq \mathbf{f}(\mathbf{y}) = \mathbf{x}^{\text{m}}(y_1,y_2) + \left.\frac{\mathbf{f}^{\text{m}}_{,1} \times \mathbf{f}^{\text{m}}_{,2}}{|\mathbf{f}^{\text{m}}_{,1} \times \mathbf{f}^{\text{m}}_{,2}|}\right|_{(y_1,y_2)} y_3,
\end{equation}
and the desired mapping function appearing in Eq.~\eqref{TDF_general} should be given by $\mathbf{y} = \mathbf{f}^{-1}(\mathbf{x})$. Normally, an actual manifold is assigned with a thickness, say $h_0$, and this can be controlled by constraining the range of $y_3$ by letting $y_3\in(-h_0/2,h_0,2)$. Finally, one may use the treatment of function composition to define the TDF for a given manifold decorated with microstructures (generated from $\phi(y_1,y_2)=\zeta$, $(y_1,y_2)\in\Upsilon^{\text{m}}$) by
\begin{equation} \label{chi_surface}
\chi(\mathbf{x}) = \left\{
\begin{aligned}
& \phi\left(\frac{f^{-1}_1(\mathbf{x})}{h}, \frac{f^{-1}_2(\mathbf{x})}{h}\right) - \zeta(\mathbf{x}), \quad && \mathbf{f}^{-1}(\mathbf{x})\in\Upsilon^{\text{m}}\times\left(-\frac{h_0}{2}, \frac{h_0}{2}\right);\\
& -\delta, && \text{elsewhere,}
\end{aligned}\right.
\end{equation}
where $\delta$ is an arbitrary positive number; $f^{-1}_i$ denotes the $i$-th entry of $\mathbf{f}^{-1}$. As from Eq.~\eqref{chi_surface}, the TDF values for all points falling outside the thickness of the manifold is negative, and no solid materials are distributed there according to Eq.~\eqref{TDF_def}.

\begin{figure}[!ht]
  \centering
  \includegraphics[width=.9\textwidth]{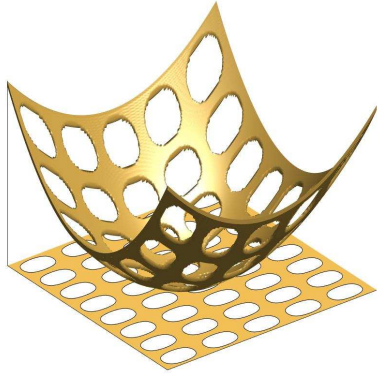}
  \caption{Infill configurations on manifolds generated based on Eqs.~\eqref{chi_surface} and \eqref{map_surface_example1}.}
  \label{Fig_infill_on_manifolds}
\end{figure}

One illustrative microstructural surface is shown in Fig.~\ref{Fig_infill_on_manifolds}.
The generating cell has a circular hole in its middle, which is given by $\phi(y_1,y_2) = \sqrt{y_1^2+y_2^2} - 0.8$ with $(y_1,y_2)\in(-0.5,0.5)^2$. The manifold is chosen as a parabolic surface given by $x_3 = x_1^2+x_2^2$. Following the aforementioned steps, the map between the fictitious and the actual spaces is found to be
\begin{equation} \label{map_surface_example1}
\begin{pmatrix} x_1\\ x_2 \\ x_3 \end{pmatrix} = \begin{pmatrix} y_1\\ y_2 \\ y_1^2+y_2^2 \end{pmatrix} + \frac1{1+4y_1^2+4y_2^2} \begin{pmatrix} -2y_1\\ -2y_2 \\ 1 \end{pmatrix}\cdot y_3.
\end{equation}
Incorporating this into Eq.~\eqref{chi_surface} and letting $h=0.2$, $h_0=0.1$, $\zeta(\mathbf{x})=0$, we get the microstructural surface in Fig.~\ref{Fig_infill_on_manifolds}.

\begin{figure}[!ht]
  \centering
  \subfigure[]{\includegraphics[width=.4\textwidth]{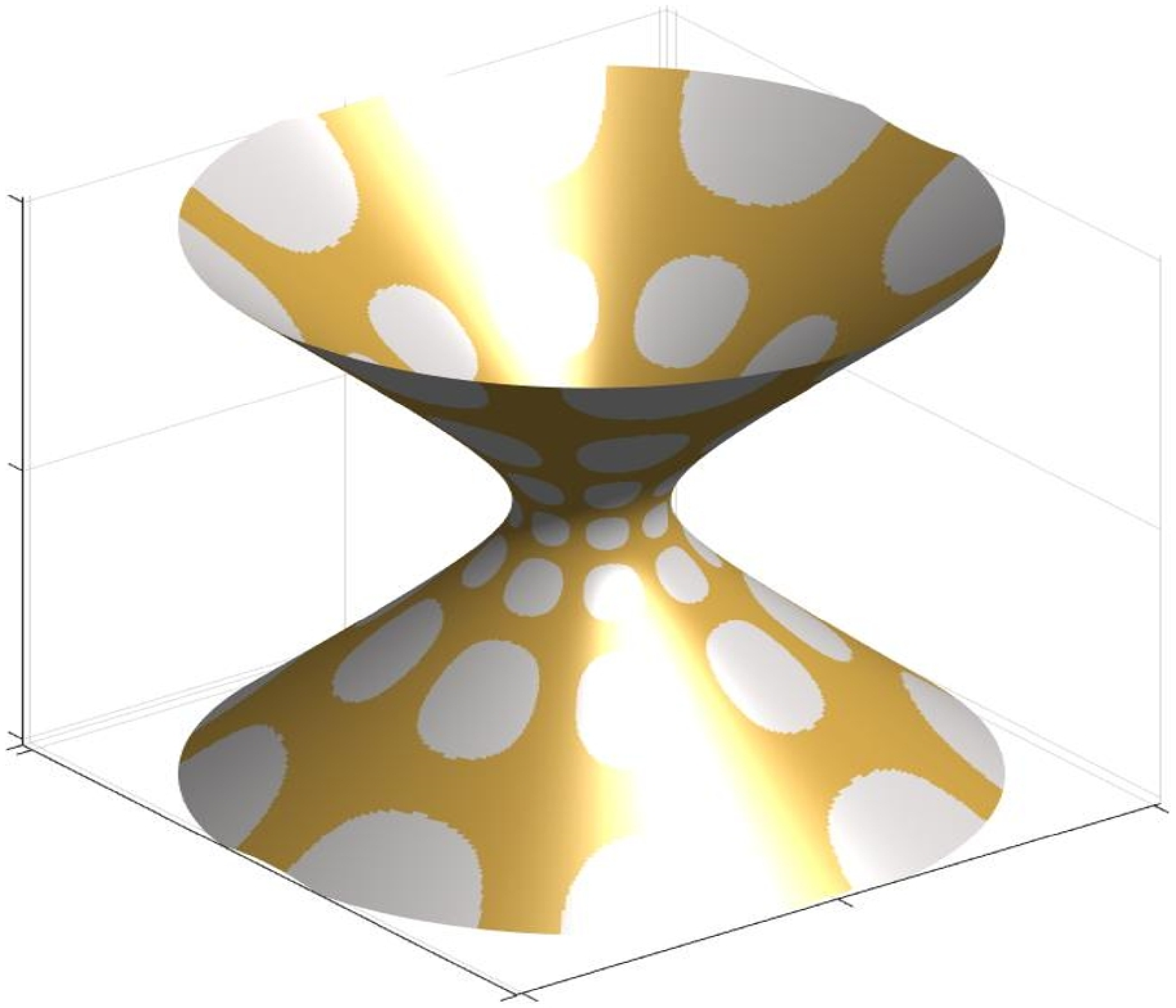}}
  \subfigure[]{\includegraphics[width=.4\textwidth]{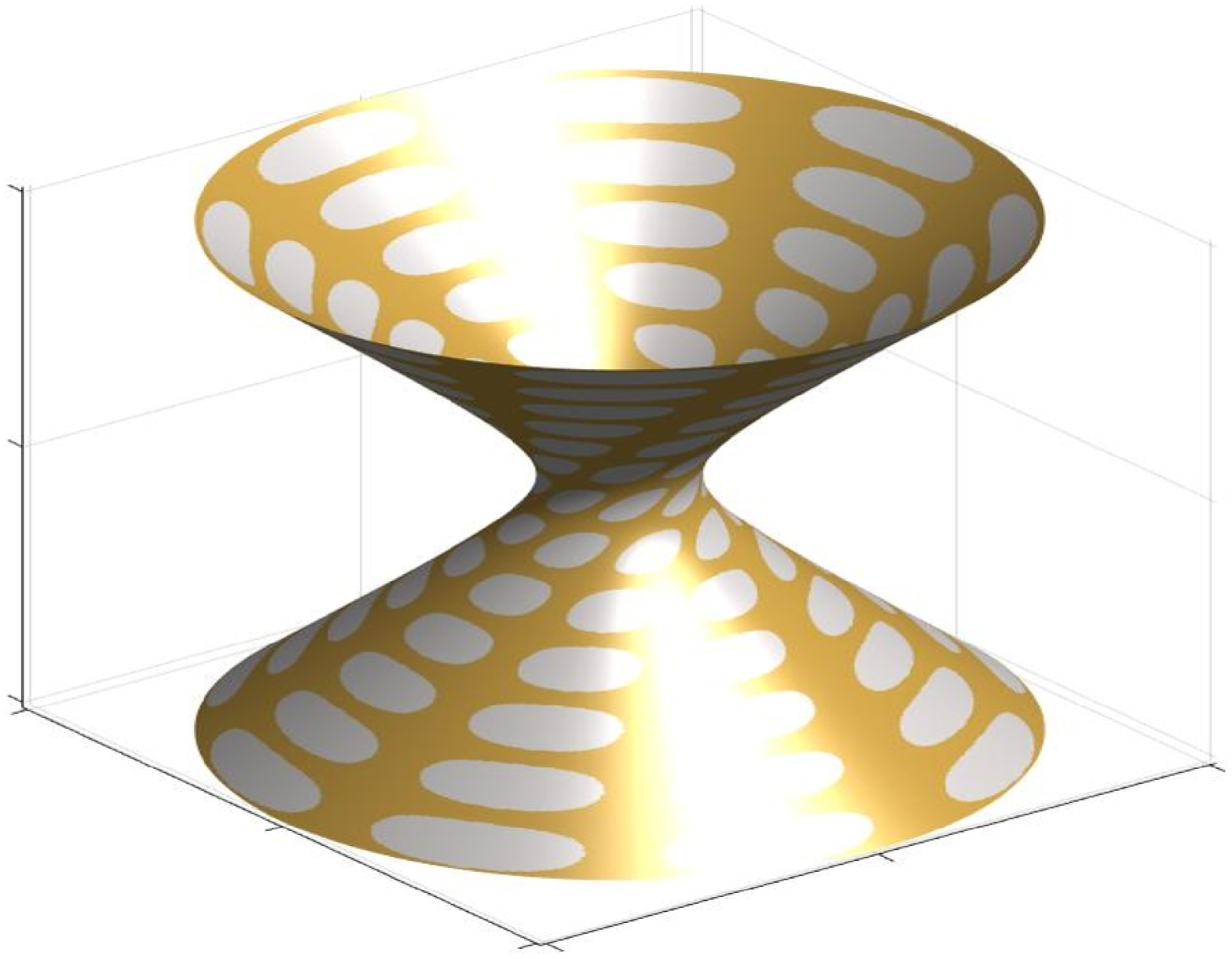}}
  \caption{Different microstructures decorated on a same hyperboloid.}
  \label{Fig_hyperboloid}
\end{figure}

It is noted that for a same manifold, different parameterisations lead to disparate microstructural surfaces. For instance, both
\begin{subequations}
\begin{equation} \label{hyperboloid1}
\{\mathbf{x}^{\text{m}}| x_1^{\text{m}} = \cosh t\cos\theta, \, x_2^{\text{m}} = \cosh t\sin\theta, \, x_3^{\text{m}} = \sinh t,\, \theta\in[0,2\pi),\, t\in\mathbb{R}\}
\end{equation}
and
\begin{equation} \label{hyperboloid2}
\{\mathbf{x}^{\text{m}}| x_1^{\text{m}} = \cos\theta - t\sin\theta, \, x_2^{\text{m}} = \sin\theta + t\cos\theta, \, x_3^{\text{m}} = t,\, \theta\in[0,2\pi),\, t\in\mathbb{R}\}
\end{equation}
\end{subequations}
correspond to a same hyperboloid. But they give rise to different manifold decorations, as shown in Fig~\ref{Fig_hyperboloid}.
Note that the hyperboloid from Eq.~\eqref{hyperboloid2} is obtained by revolving a non-vertical straight line (parameterised by $t$). Thus straight microstructural orientations are seen in Fig.~\ref{Fig_hyperboloid}(b).

\subsection{Comparison among existing methods for IGM representation\label{Sec_unification}}
The present treatment based on Eq.~\eqref{TDF_general} effectively offers a unified theoretical framework, which encompasses most existing approaches for IGM representation. Actually, Eq.~\eqref{TDF_general} (and its extended form by Eq.~\eqref{chi_interpolation}) enables two sets of operations: selection of generating cell configurations by the value of $\zeta(\mathbf{x})$, and the manipulation of the selected cell (involving cell rescaling, cell rotation and cell distortion in an iso-volumetric manner) controlled by the mapping function $\mathbf{y}(\mathbf{x})$, or alternatively, by the corresponding Jacobian matrix $\mathbf{J}$ given by
\begin{equation} \label{Jacobian}
J_{ij} = \frac{\partial y_i}{\partial x_j}.
\end{equation}
In theory, $\mathbf{J}$ controls the operation exerted on a single cell, and $\zeta(\mathbf{x})$ controls the transition between different cells. In this viewpoint, most existing treatments for IGM generation can be considered as being carried out through varying $\mathbf{J}$ and $\zeta(\mathbf{x})$, and a summary is made in Table~\ref{Table_comparison}.
\begin{table}[!ht]
  \centering
  \begin{tabular}{c|c|c|c|c}
  & & & Smooth & \\
  & Cell shape & $\mathbf{J}$ & -ness & $\zeta(\mathbf{x})$\\
  \hline
  Classical & Oriented & & & Piece-wise\\
  HBMs & cuboid & $\mathbf{D}$ & No & constant\\
  \hline
   & Oriented & & Slightly & Discrete\\
  MPCs & cuboid & $\mathbf{D}$ & not & step function\\
  \hline
   & Oriented & & Slightly & Discrete\\
  MLSMs & cuboid & $\mathbf{D}$ & not & step function\\
  \hline
  De & Rotational & & & \\
  -homogenisation & cuboid & $\mathbf{D}\mathbf{Q}$ & Yes & Continuous\\
  \hline
  Original & & & & \\
  AHTO plus & parallelepiped & $\det\mathbf{J}\neq0$ & Yes & Contant\\
  \hline
  \end{tabular}
  \caption{A summary of existing treatments for IGM configurations under the framework given by Eq.~\eqref{TDF_general}. $\mathbf{D}$ denotes a diagonal matrix, and $\mathbf{Q}$ denotes an orthogonal matrix.\label{Table_comparison}}
\end{table}

Here the reported IGM generation approaches in literature are in general classified into five groups:
\begin{enumerate}
  \item[1)] classical homogenisation-based methods (HBMs) \cite{Bendsoe_CMAME1988,Bendsoe_StructOpt1989,Rodrigues_SMO2002,Coelho_SMO2008,NiuB_SMO2009} where each pixel is associated with a microstructural cell;
  \item[2)] methods with parameterised cells (MPCs) \cite{Radman_JMaterSci2013,WangC_SMO2018,ChengL_CMAME2019,LiQH_CompStruct2019,WangC_SMO2020} where the constituting cells are formed by parameterised building blocks;
  \item[3)] multiple level-set methods (MLSMs) \cite{WangYQ_CMAME2017, ZhangY_SMO2019} where the constituting cells are identified by multiple level-set contours of discrete values from a same generating function;
  \item[4)] de-homogenisation methods \cite{Groen_IJNME2018,Allaire_CompMathAppl2019};
  \item[5)] the original AHTO plus method \cite{ZhuYC_JMPS2019,XueDC_arxiv2019}.
\end{enumerate}

For the first three groups of methods, the constituting cells must take a rectangle/cuboid shape. In a viewpoint of Eq.~\eqref{TDF_general}, each $y_i$ is only a function $x_i$, for $i=1$, $\cdots$, $n$, or alternatively, the associated Jacobian matrix $\mathbf{J}$ must be diagonal, i.e., $\mathbf{J}=\mathbf{D}$, with $\mathbf{D}$ a diagonal matrix. In contrast, with the projection algorithm \cite{Pantz_SIAMJCO2008} used in de-homogenisation methods, cell rotation is also permitted. In the context of Eq.~\eqref{TDF_general}, $\mathbf{J}=\mathbf{D}\mathbf{Q}$ with $\mathbf{Q}$ an orthogonal matrix. And for the original AHTO plus method, the choice of $\mathbf{J}$ is more arbitrary provided that $\det\mathbf{J}\neq0$. But in theory, a pool with $\mathbf{J}=\mathbf{D}\mathbf{Q}$ should suffice for optimising the IGM mechanical behaviour, and this is well demonstrated by Groen et al. \cite{Groen_IJNME2018}. It should also be noted that, the edges of the constituting cells, in categories 4) and 5), are slightly curved in the actual space so as to accommodate smooth connection across cell boundaries.

The IGM smoothness is also controlled by the continuity in function $\zeta(\mathbf{x})$ appearing in Eq.~\eqref{TDF_general}. For 1) classical homogenisation-based approaches, each pixel is individually assigned with a microstructure. Thus $\zeta(\mathbf{x})$ is effectively piece-wisely defined over individual pixels. Therefore, non-smoothness emerges in the resulting IGM configurations. For category 2) MPCs, the microstructure is generated through gradually varying the parameter values of its building blocks, but $\zeta(\mathbf{x})$ stays the same in every single cell. Hence $\zeta(\mathbf{x})$ should be a step function, where the gaps between neighbouring steps are tiny. Consequently, the resulting IGM configurations appear gradually-varying, but slightly non-smooth connection still emerges across cell boundaries. Similar case arises for the category 3) MLSMs, where the microstructure is generated through gradually controlling the outputs of a single level-set function. In contrast, the de-homogenisation methods enable smooth transition across cell boundaries in the actual space, which should correspond to a continuous function $\zeta(\mathbf{x})$. As for the original AHTO plus method, no transition between cells is permitted, giving rise to a constant-value function of $\zeta(\mathbf{x})$. But this limitation is removed in the present article by Eq.~\eqref{TDF_general}.

Therefore, most existing methods for IGM generation should find their corresponding formulation through Eq.~\eqref{TDF_general} by selecting appropriate $\mathbf{J}$ and $\zeta(\mathbf{x})$.

\section{Asymptotic formulation towards optimising IGM compliance\label{Sec_homogenisation}}
In this section, we will first consider how to formulate the mechanical behaviour of IGM configurations obtained based on Eq.~\eqref{TDF_general}. Then an optimisation problem for IGM compliance is established, with the corresponding sensitivity analysis being discussed at the end of the section.

\subsection{IGM mechanical behaviour}
Under the original AHTO plus framework \cite{ZhuYC_JMPS2019}, the multiscale mechanical response of a loaded IGM configuration can be predicted through asymptotic analysis. This is carried out by approximating the (multiscale) displacement field by a series expanded in terms of the small parameter $\epsilon$ defined by Eq.~\eqref{scale_separation}. Each term in the asymptotic expansion is a function containing two sets of spatial variables, say $\mathbf{x}$ and $\bar{\mathbf{Y}}$. The variable $\mathbf{x}$ captures the homogenised behaviour as if the IGM configuration is envisaged as a continuum. The (non-dimensional) variable $\bar{\mathbf{Y}}=\mathbf{y}(\mathbf{x})/h$ captures the large displacement gradient accommodating the local microstructures. Through asymptotic analysis, the originally multiple-scale governing equations are solved on different scales individually.

In the following, the key equations constituting the original AHTO plus formulation \cite{ZhuYC_JMPS2019} are outlined. A weak form of the equilibrium equation is set up for the homogenised displacement field $\mathbf{u}^{\mathrm{H}}$, given by
\begin{subequations}\label{eqn_homogenised}
\begin{equation} \label{eqn_fb_homogenised}
\int_\Omega \mathbb{C}_{ijkl}^\mathrm{H}u_{i,j}v_{k,l}\d \mathbf{x} = \int_\Omega f_i^\mathrm{H}v_i \d \mathbf{x} + \int_{\Gamma_{\text{t}}}\tilde{t}_iv_i\d S,\qquad \forall\,\mathbf v \in \mathcal U_{\text{ad}},
\end{equation}
where $\mathbb{C}^\mathrm{H} = (\mathbb{C}^\mathrm{H}_{ijkl})$ denotes the effective elasticity tensor of the homogenised continuum; $\mathbf v$ is a virtual displacement field belonging to the functional space of $\mathcal U_{\text{ad}} = \{\mathbf v| \left.\mathbf v\right|_{\Gamma_{\text{u}}}=0\}$; $\Gamma_{\text{u}}$ denotes the boundary on $\Omega$ where displacement boundary condition is imposed; $\Gamma_{\text{t}}$ is the boundary on $\Omega$ where a surface traction of density $\tilde{\mathbf{t}}$ is applied.

Meanwhile, a set of characteristic functions  $\boldsymbol{\xi}(\bar{\mathbf{Y}};\mathbf{x}) = (\xi_i^{jk}(\bar{\mathbf{Y}};\mathbf{x}))$, $i$, $j$, $k=1$, $\cdots$, $n$, are also needed, and they are obtained by solving the following set of PDEs:
\begin{equation} \label{eqn_cell_homogenised}
J_{mj}\pd{}{\bar Y_m}\left(\tilde{\mathbb{C}}_{ijkl}J_{nl}\pd{\xi_k^{st}}{\bar Y_n}\right)= J_{mj}\pd{\tilde{\mathbb{C}}_{ijst}}{\bar Y_m},\quad \bar{\mathbf{Y}}\in \Upsilon_\text{p}= \left[-\frac1{2},\frac1{2}\right]^n,
\end{equation}
subjected to periodic boundary conditions over $\Upsilon_{\text{p}}$, where $\mathbf{J}$ is the Jacobian matrix defined by Eq.~\eqref{Jacobian}; $\tilde{\mathbb{C}}$ is a piecewisely constant function: it equals the corresponding elasticity tensor $\mathbb{C}$ at a solid point, and vanishes elsewhere in $\Upsilon_\text{p}$. It is noted that the spatial variables of $\xi_i^{jk}(\bar{\mathbf{Y}};\mathbf{x})$ are $\bar{\mathbf{Y}}$, while $\mathbf{x}$ simply serves as a parameter. But this means one has to solve Eq.~\eqref{eqn_cell_homogenised} individually at every single macroscopic point in the actual space.

When the solutions to Eq.~\eqref{eqn_cell_homogenised} become available, one may finally evaluate the effective elasticity tensor needed for solving Eq.~\eqref{eqn_fb_homogenised}:
\begin{equation} \label{eqn_C}
\mathbb{C}_{ijkl}^\mathrm{H}=\mathbb{C}_{ijkl}\left|\Upsilon^\text{s}_\text{p}\right| -\mathbb{C}_{ijst}J_{nt}\int_{\Upsilon^\text{s}_\text{p}}\pd{\xi^{kl}_s}{\bar Y_n}\d \bar{\mathbf{Y}},
\end{equation}
where $\Upsilon^\text{s}_\text{p}$ is recalled to be the solid part in $\Upsilon_\text{p}$.
\end{subequations}

Such a scale-separation treatment roots in the fact that an IGM configuration from the original AHTO plus framework becomes periodic in the fictitious space measured by $\mathbf{y} = \mathbf{y}(\mathbf{x})$, or equivalently by $\bar{\mathbf{Y}} = \mathbf{y}/h$. For the present case, however, the generated IGM configurations are not exactly periodic in the non-dimensional fictitious space measured by $\bar{\mathbf{Y}}$. Here we still adopt the same asymptotic formulae as given by Eqs.~\eqref{eqn_fb_homogenised} - \eqref{eqn_C}. The accuracy of the treatment here will be demonstrated through a comparison with the corresponding fine-scale simulation results in Sec.~\ref{Sec_numerical_examples}.

\subsection{Topology optimisation}
In this subsection, based on the results above, a mathematical formulation for optimising IGM compliance can be set up as follows
\begin{equation} \label{topology_optimisation}
\begin{aligned}
&\text{Find}\qquad \mathbf y=\mathbf y(\mathbf x) \in C^0, \, \zeta(\mathbf{x})\in C^0, \chi^{\text{p}}(\cdot) \in \mathcal{U}\left(\Upsilon_{\text{p}}\right) \\
&\text{Minimise}\qquad \mathcal{C}^\mathrm{H}=\int_\Omega \mathbb{C}_{ijkl}^\mathrm{H}\pd{u_i^\mathrm{H}}{x_j}\pd{u_k^\mathrm{H}}{x_l}\d \mathbf{x}\\
&\text{Subject to}\\
&\qquad\text{Eqs.\eqref{eqn_fb_homogenised} - \eqref{eqn_C},}\quad \int_{\Omega} g(\zeta(\mathbf x))\, \d \mathbf x \leq \frac{V_0}{|\Omega|},
\end{aligned}
\end{equation}
where $C^0$ denotes the functional space of continuous functions; $\mathcal{U}\left(\Upsilon_{\text{p}}\right)$ is a certain functional space formed by periodic functions defined over the non-dimensional unit cell $\Upsilon_{\text{p}}$; $\mathcal{C}^\mathrm{H}$ is the homogenised structural compliance; $V_0$ is the upper limit of the IGM solid volume. The last inequality in Eq.~\eqref{topology_optimisation}, as from Eq.~\eqref{vol_fraction}, formulates the volume fraction constraint.

Eq.~\eqref{topology_optimisation} establishes a scheme for minimising IGM compliance. Compared to the original AHTO plus framework, it exhibits several advantageous features. Firstly, the available IGM configurations are far more abundant. This is because one can not only conduct IGM design by rotating and distorting a generating cell through the mapping function $\mathbf{y}(\mathbf{x})$, but also select among generating cells through the level-set indicator function $\zeta(\mathbf{x})$. Secondly, under the original AHTO plus framework, one can not increase the volume of solid materials to strengthen the local microstructure, because the volume fraction of all constituting cells must be identical to each other. But this limitation is fully removed now. It will be shown in Sec.~\ref{Sec_numerical_examples} that enabling transition among different transition cells should bring down the compliance value of an optimised structure by almost a half.

\subsection{Gradient-based sensitivity analysis\label{Sec_sensitivity}}
For speeding up the optimisation process based on Eq.~\eqref{topology_optimisation}, its associated sensitivity analysis has been performed. Suppose that the vector $\mathbf d=(d_1,\dots,d_\lambda)^\top$ quantifies all the design variables. Through using the idea of the adjoint method \cite{LiuST_SMO2002}, the derivative of the IGM compliance with respect to the $\tau$-th design variable can be formulated by
\beq\label{sensitivity_C}
\pd{\mathcal{C}^\mathrm{H}}{d_\tau}=-\int_{\Omega}\pd{ \mathbb{C}^\mathrm{H}_{ijkl}}{d_\tau}\pd{u^\mathrm{H}_i}{x_j}\pd{
u^\mathrm{H}_k}{x_l}\,\d \mathbf{x}.
\eeq
Thus one simply needs to compute the derivative of the elasticity tensor $\mathbb{C}^{\mathrm{H}}$ with respect to all the design variables. Here the design variables are classified into two groups: the macroscopic design variables parameterising the mapping function $\mathbf{y}(\mathbf{x})$ and the level-set indicator function $\zeta(\mathbf{x})$; the microscopic design variables parameterising the structural generating function $\phi(\bar{\mathbf{Y}})$.

For sensitivity analysis with respect to typical macroscale variables, say, $d_{\alpha}$, we follow \cite{XueDC_arxiv2019} to write down
\beq \label{sensitivity_macro}
\pd{\mathbb{C}^\mathrm{H}_{ijkl}}{d_\alpha}=\pd{\mathbb{C}^{\mathrm{H}[1]}_{ijkl}}{d_\alpha} + \pd{ \mathbb{C}^{\mathrm{H}[2]}_{ijkl}}{d_\alpha}
\eeq
where
\beq
\pd{\mathbb{C}^{\mathrm{H}[1]}_{ijkl}}{d_\alpha} = -\int_{\Upsilon_\text{p}} \left[\pd{J_{nq}}{d_\phi}\pd{\xi^{ij}_p}{\bar Y_n}\tilde{\mathbb{C}}_{pqkl} + \pd{J_{mt}}{d_\phi}\pd{\xi^{kl}_s}{\bar Y_m}\tilde{\mathbb{C}}_{ijst} \right]\d {\bar{\mathbf{Y}}}
\eeq
and
\beq
\pd{\mathbb{C}^{\mathrm{H}[2]}_{ijkl}}{d_\alpha} = \int_{\Upsilon_\text{p}} \left[\pd{J_{nq}}{d_\alpha}\pd{\xi^{ij}_p}{\bar Y_n}\left(\tilde{\mathbb{C}}_{pqst}J_{mt} \pd{\xi^{kl}_s}{\bar Y_m}\right) + \pd{J_{mt}}{d_\alpha}\pd{\xi^{kl}_s}{\bar Y_m}\left(\tilde{\mathbb{C}}_{pqst}J_{nq}\pd{\xi^{ij}_p}{\bar Y_n}\right) \right]\d {\bar{\mathbf{Y}}},
\eeq
respectively.

As for the sensitivity with respect to the microscale variables, say, $d_{\beta}$, we also follow \cite{XueDC_arxiv2019} to write down
\beq \label{sensitivity_micro}
\begin{aligned}
\pd{\mathbb{C}^{\mathrm{H}}_{ijkl}}{d_{\be}} &= \int_{\Upsilon_\text{p}}\left(\delta_{ip}\delta_{jq}-J_{nq}\pd{\xi^{ij}_p}{\bar Y_n}\right) \pd{\tilde{\mathbb{C}}_{pqst}}{d_{\beta}} \left(\delta_{ks}\delta_{lt}-J_{mt}\pd{\xi^{kl}_s}{\bar Y_m}\right) \, \d {\bar{\mathbf{Y}}},
\end{aligned}
\eeq
where $\delta_{ij}$, $i$, $j=1$, $\cdots$, $n$, is the Kronecker delta.

\section{Numerical results and analysis\label{Sec_numerical_examples}}
In this section, numerical results based on the proposed scheme are presented. Two-dimensional examples will first be examined in depth, so as to reveal the key features of the present method. Then three-dimensional numerical examples will also be briefly studied at the end of the section.

\subsection{General problem setting}
For all the examples examined in this section, linearly elastic material response is assumed with the isotropic parameters chosen as follows: the (nondimensional) Young's modulus $E=1$ and Poisson's ratio $\nu=0.3$. For two-dimensional examples, the specimen's (non-dimensional) thickness in the third dimension is fixed to be 1, and scenarios of plane stress are considered. The upper limit of IGM volume fraction is set to be 0.3 unless otherwise specified.

Here the continuous menu of generating cell configurations, once chosen, are set fixed during the optimisation processes. The reason for this action is that IGM generation based on Eq.~\eqref{TDF_general} should output abundant configurations. This is in particular reasonable, when IGM configurations are produced through continuously patching seminal cell structures. Thus the evolution of the design variables parameterising the structural generating function $\phi(\cdot)$ is temporarily suspended here. As for the parameterisation of the mapping function $\mathbf{y}(\mathbf{x})$, we follow \cite{XueDC_arxiv2019} to assume
\beq\label{y_poly}
y_{i}=a_{ij}x_{j}+\frac{1}{2}b_{ijk}x_{j}x_{k}+\frac{1}{3}c_{ijkl}x_{j}x_{k}x_{l}, \quad i, j, k, l = 1, \cdots, n,
\eeq
where symmetric conditions $c_{ijkl}=c_{ijlk}=c_{ilkj}=c_{ikjl}$ and $b_{ijk}=b_{ikj}$ are imposed.

As for the level-set indicator function $\zeta(\mathbf{x})$, we let
\begin{equation} \label{zeta_poly}
\zeta(\mathbf{x}) = \alpha + \beta_k x_k + \frac{1}{2} \gamma_{kl} x_kx_l,
\end{equation}
where $\gamma_{kl} = \gamma_{lk}$ for $k$, $l=1$, $\cdots$, $n$, is required.

Upon this set-up, the design variables are collected by
\begin{equation} \label{design_variabls_poly}
\{a_{ij}, b_{ijk}, c_{ijkl}; \alpha, \beta_{i}, \gamma_{ij}\}_{i,j,k,l=1}^n.
\end{equation}
Hence we have 24 design variables for two-dimensional problems, and 67 design variables for three-dimensional cases. Here the length-scale characterising parameter $h$ is set to be 0.05, and the method of moving asymptotes (MMA) \cite{Svanberg_IJNME1987} are employed as the optimiser in this work.

Note that the homogenisation formulation for stress analysis involves individually solving a cell problem defined by Eq.~\eqref{eqn_cell_homogenised} at every single macroscopic point $\mathbf{x}$. This significantly brings up the computational cost. For overcoming this issue, a zoning scheme \cite{XueDC_arxiv2019} is employed, where the design domain is partitioned into several subdomains. In each subdomain, a point is selected and the effective elasticity tensor at that point obtained through Eq.~\eqref{eqn_cell_homogenised} is used to represent that of the whole subdomain. Here all design domains are divided into rectangles/cuboids, and the center of each rectangle/cuboild is chosen as the representative point. It has been shown \cite{XueDC_arxiv2019} that, when a subdomain contains no more than 25 finite elements (for two-dimensional cases), the deviation from the corresponding fine-scale results is less than 0.04. We will resume using this criterion for conducting simulations for this work. As for finite element analysis, four-node quadrilateral elements are used in all two-dimensional examples while eight-node hexahedral brick elements are adopted in all three-dimensional examples.

\subsection{Two-dimensional cases}
\subsubsection{Laminar IGM configurations\label{Sec_case2D1}}
\begin{figure}[!ht]
  \centering
  \includegraphics[width=.6\textwidth]{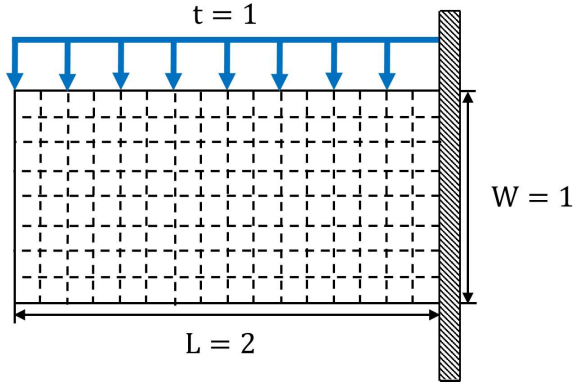}
  \caption{A short beam subjected to uniformly distributed force.}
  \label{short_beam_2D_uniform}
\end{figure}

We start with an example of a short beam fixed on its right side, as shown in Fig.~\ref{short_beam_2D_uniform}.
A uniformly distributed compression of magnitude 0.1 is applied over its top side. The design domain is of size $2\times1$, and it is divided into $16\times8$ identical square-shape subdomains. A $400\times200$ mesh is used for solving the homogenised problem.

\begin{figure}[!ht]
  \centering
  \includegraphics[width=.75\textwidth]{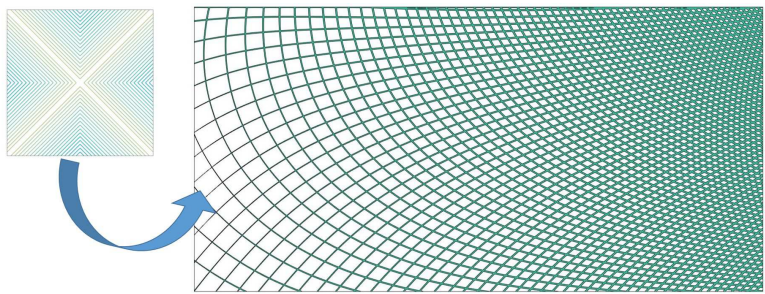}
  \caption{The optimised design obtained with X-shape material distribution in the unit generating cell.}
  \label{uni_X_2D}
\end{figure}

The continuous menu of generating cells are selected as a set of ``X''-shape configurations, which are controlled by the thickness of the solid region. Mathematically, all generating cells can be represented by the multiple level-set contours of a single function:
\begin{equation} \label{fun_generation_X}
\phi(\bar{\mathbf{Y}}) = \frac{\sqrt{2}}{4} - \left||\bar{Y}_1| - |\bar{Y}_2|\right|, \quad \bar{\mathbf{Y}} \in \Upsilon_{\text{p}} = \left[-\frac1{2}, \frac1{2}\right]^2,
\end{equation}
and the structural boundaries of several generating cell configurations are shown in the left panel of Fig.~\ref{uni_X_2D}.

Then according to Eq.~\eqref{TDF_general}, the TDF associated with the corresponding IGM configuration is thus given by
\begin{equation} \label{TDF_X}
\chi(\mathbf{x}) = \phi\left(\frac{\mathbf{y}(\mathbf{x})}{h}\right) - \zeta(\mathbf{x}),
\end{equation}
where $\mathbf{y}(\mathbf{x})$ and $\zeta(\mathbf{x})$ are given by Eqs.~\eqref{y_poly} and \eqref{zeta_poly}, respectively.

Note that the range of $\phi(\cdot)$ by Eq.~\eqref{fun_generation_X} defines the bounds for $\zeta(\mathbf{x})$, that is,  $\zeta(\mathbf{x})\in [0,\sqrt{2}/4]$. In practice, if an output of Eq.~\eqref{zeta_poly} exceeds $\sqrt{2}/4$, we require it equal $\sqrt{2}/4$, and so does an output falling below $0$. Under the present setting, the overall IGM volume fraction, according to Eq.~\eqref{vol_fraction}, is formulated by
\beq\label{volume_X}
\frac{V}{|\Omega|} = 1 - \frac8{|\Omega|} \int_{\Omega}\left(\zeta(\mathbf{x}) \right)^2 \, \d \mathbf{x}.
\eeq

\begin{figure}[!ht]
  \centering
  \includegraphics[width=.7\textwidth]{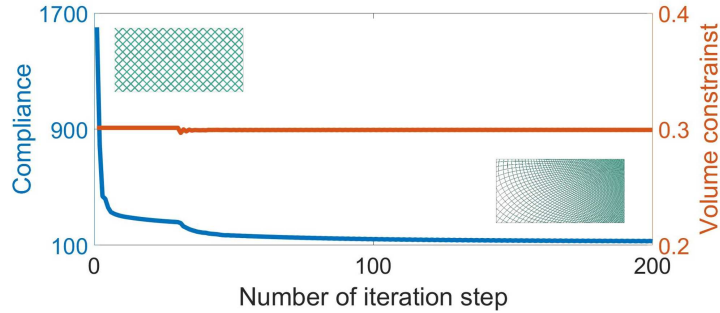}
  \caption{Convergence curve of the IGM compliance.}
  \label{curve_converge_compliance}
\end{figure}

The optimised result is then shown in the right panel of Fig.~\ref{uni_X_2D}, and the IGM volume fraction is roughly 29.98\%. The structural compliance based on the homogenisation formulation is computed to be 126.60, which delivers a 2.0\% deviation from the corresponding fine-scale simulation result (124.24). Fig.~\ref{curve_converge_compliance} shows the evolution of both the compliance value and the corresponding volume fraction during optimisation. Compared to the initial structure composed of a spatially periodic microstructure, the structural compliance value drops more than 90\%. It is noted that the start of evolution of the level-set indicator function $\zeta(\mathbf{x})$ is 30 steps later than that of the mapping function $\mathbf{y}(\mathbf{x})$. This is because for minimising structural compliance, rotation and distortion of a single cell seems more effective than simply transiting among generating cells. This will be discussed with greater details later.

Compared to the spatially periodic configuration, the optimised one arranges the constituting fibres such that the external load on the top side of the design domain smoothly transited to the fixed boundary. Besides, the material distribution is denser at the top-right corner, where more materials are needed for accommodating the relatively high local bending moment.

\begin{figure}[!ht]
  \centering
  \includegraphics[width=0.7\textwidth]{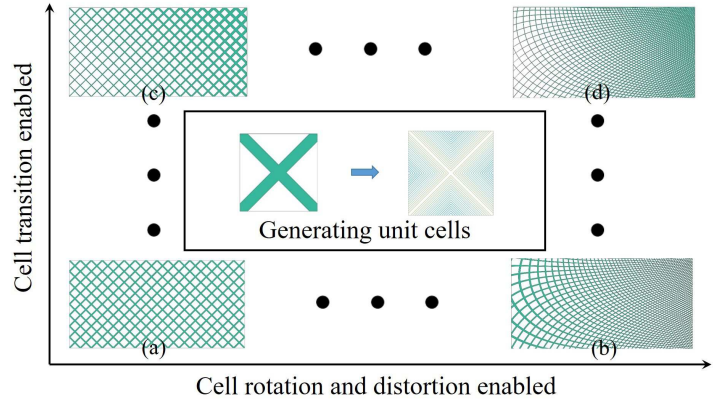}
  \caption{Different means for IGM generation based on the continuous menu of generating cells given in the left panel of Fig.~\ref{uni_X_2D}.\label{figure_x}}
\end{figure}

Eq.~\eqref{TDF_general} effectively generates an IGM configuration combinatively by two means. One is through distorting and rotating a single cell by the mapping function $\mathbf{y}(\mathbf{x})$, and the other is through transition among generating cells by the level-set indicator $\zeta(\mathbf{x})$. Their individual roles in compliance optimisation are examined with the results summarised in Fig.~\ref{figure_x} and Table~\ref{Table_X}.
\begin{table}[!ht]
  \centering
  \begin{tabular}{c|c|c|c|c}
  & Periodic& Cell manipulation & Cell transition & Both\\
  & structure& enabled only& enabled only & enabled\\
  \hline
   & & & &\\
  Compliance & 1601.91 & 205.15 & 742.61 & 126.60\\
  \hline
  Volume& & & &\\
  fraction (\%) & 30.00 & 30.00 & 29.99 & 29.93\\
  \hline
  Optimised& & & &\\
  design & Fig.~\ref{figure_x}(a) & Fig.~\ref{figure_x}(b) & Fig.~\ref{figure_x}(c) & Fig.~\ref{figure_x}(d)\\
  \hline
  \end{tabular}
  \caption{Roles played by manipulation of a single cell and transition among generating cells. Here cell manipulation involves cell rotation and distortion.\label{Table_X}}
\end{table}

It is shown that optimisation by simply permitting rotation and distortion of a single cell reduces the compliance value to 1/8 of that of the periodic structure shown in Fig.~\ref{figure_x}(a). This is in contrast to a drop of 1/2 if only transition between generating cells is enabled. This means arranging laminar orientations in align with the local directions of principal stress provides a more effective mean than simply redistributing materials by means of cell transition. However, compared to the optimised configuration in Fig.~\ref{figure_x}(b), further enabling transition among the generating cells results in an extra 50\% drop in the optimised compliance. This demonstrates the necessities of involving transition among different cell configurations during IGM generation.

\subsubsection{IGM generated through interpolating between seminal cells}
\begin{figure}[!ht]
  \centering
  \includegraphics[width=.75\textwidth]{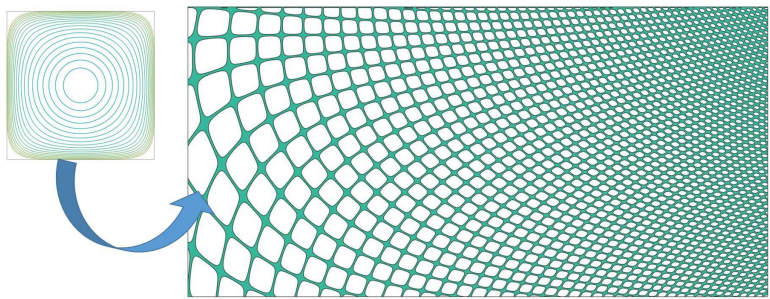}
  \caption{An IGM design optimised out of a continuous menu of generating cells obtained through interpolating between a full solid and a cell with a hyperelliptic hole in its middle.}
  \label{uni_C_2D}
\end{figure}

In this case, the continuous menu of generating cells are obtained by interpolating between two seminal cells, which are chosen as a full solid square with a diminishing circular hole at its centre and a unit square with a hyperelliptic hole in its middle. Mathematically, a full-solid configuration with a diminishing hole can be identified by the zero level-set contour of
\begin{equation} \label{case2D2_cell1}
\phi_1(\bar{\mathbf{Y}}) = \bar{Y}_1^2 + \bar{Y}_2^2, \quad \bar{\mathbf{Y}} \in \Upsilon_{\text{p}}^{\text{s}} = \left[-\frac1{2},\frac1{2}\right]^2.
\end{equation}
The cell with a hyperelliptic hole at its centre is indicated by the zero level-set contour of
\begin{equation} \label{case2D2_cell2}
\phi_2(\bar{\mathbf{Y}}) = \bar{Y}_1^6 + \bar{Y}_2^6 - \frac1{64}, \quad \bar{\mathbf{Y}} \in \left[-\frac1{2},\frac1{2}\right]^2.
\end{equation}
Thus based on Eq.~\eqref{phi_interpolation}, one introduces a structural generating function $\phi(\cdot)$ by interpolating $\phi_1$ and $\phi_2$ controlled by a parameter $\zeta\in[0,1]$:
\begin{equation} \label{case2D2_chip}
\phi(\bar{\mathbf{Y}}) = (1-\zeta)\left(\bar{Y}_1^2 + \bar{Y}_2^2\right) + \zeta\left(\bar{Y}_1^6 + \bar{Y}_2^6\right) + \frac{63\zeta}{64},
\end{equation}
such that the implicit relation $\phi(\bar{\mathbf{Y}}) = \zeta$ corresponds to the structural boundary of one generating cell configuration in the fictitious space measured by $\bar{\mathbf{Y}}$. The structural boundaries of several generating cells are shown in the left panel of Fig.~\ref{uni_C_2D}.
It can be checked that when $\zeta$ is near 0, $\phi_1$ casts a dominant weight over $\phi_2$. In this scenario, $\phi(\bar{\mathbf{Y}}) = \zeta$ identifies a single point, and this corresponds to a square of full solid. As $\zeta$ keeps growing, the hole shape looks more like a mixture of a circle and a hyper-ellipse. When $\zeta=1$, $\phi(\bar{\mathbf{Y}})=\zeta$ becomes $\phi_2(\bar{\mathbf{Y}}) = 0$, corresponding to a cell with a hyperelliptic hole. Note that the transition in the hole shape is carried out smoothly, since $\phi(\cdot)$ given by Eq.~\eqref{case2D2_chip} is a continuous function (in $\zeta$).

The volume fraction of the generating cells, obtained based on Eq.~\eqref{case2D2_chip}, spans from 100\% ($\zeta=0$) to 3.81\% ($\zeta=1$). In contrast to the previous case, no explicit expressions linking a generating cell with its corresponding volume fraction are available. Nevertheless, one may pre-calculate a data set correlating the two quantities, and fit for a simple function approximating the volume fraction of generating cells against the parameter $\zeta$ by
\begin{equation} \label{case2D2_volume_lambda}
g(\zeta) = (1-\zeta)^{1.35}.
\end{equation}
Finally, by introducing the mapping function by $\mathbf{y}(\mathbf{x}) = h\bar{\mathbf{Y}}$, and the level-set indicator function $\zeta(\mathbf{x}) = \zeta$, we express the TDF of the IGM configurations for topology optimisation:
\begin{equation} \label{case2D2_chi}
\begin{aligned}
\chi(\mathbf{x}) & = \left(1-\zeta(\mathbf{x})\right) \left(\left(\frac{y_1(\mathbf{x})}{h}\right)^2 + \left(\frac{y_2(\mathbf{x})}{h}\right)^2\right) \\ &\quad + \zeta(\mathbf{x}) \left(\left(\frac{y_1(\mathbf{x})}{h}\right)^6 + \left(\frac{y_2(\mathbf{x})}{h}\right)^6 - \frac1{64}\right),
\end{aligned}
\end{equation}
which is effectively Eq.~\eqref{chi_interpolation}. The volume constraint, based on \eqref{vol_fraction} and \eqref{case2D2_volume_lambda}, is thus formulated by
\begin{equation} \label{case2D2_volume}
\int_{\Omega} \left(1-\zeta(\mathbf{x})\right)^{1.35} \, \d \mathbf{x} \le 0.3.
\end{equation}

\begin{figure}[!ht]
  \centering
  \includegraphics[width=.7\textwidth]{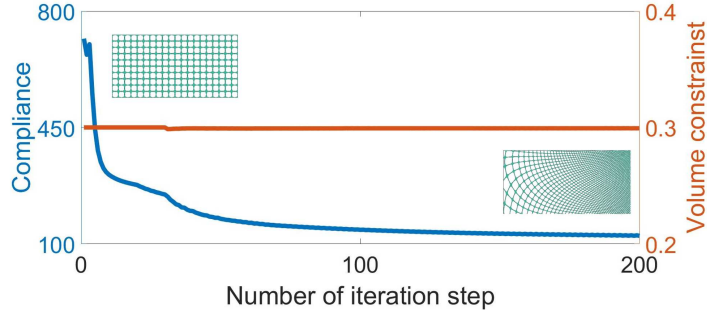}
  \caption{An optimised GMC design for the short beam example. The corresponding continuous menu of generating cells are given by a structural generating function defined by Eq.~\eqref{case2D2_chip}}
  \label{curve_converge_compliance_c}
\end{figure}

The corresponding optimised IGM design is shown in the right panel of Fig.~\ref{uni_C_2D} with the volume fraction being 29.94\%. The homogenised result for structural compliance is 124.40, which delivers a deviation of 2.1\% from the fine-scale computation (121.73). Fig.~\ref{curve_converge_compliance_c} shows the evolution in the values of both the structural compliance and the corresponding volume fraction during optimisation.

\begin{figure}[!ht]
\centering
\includegraphics[width=0.7\textwidth]{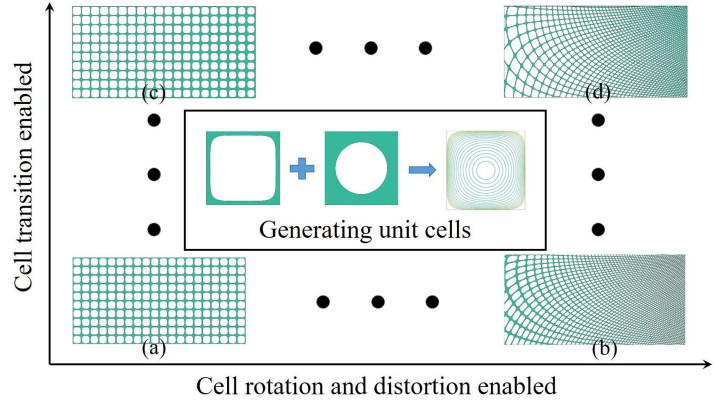}
\caption{Optimised design with different choices of unit generating cells.}
\label{figure_c}
\end{figure}

Again, the individual effects of cell manipulation (through $\mathbf{y}(\mathbf{x})$) and cell transition (through $\zeta(\mathbf{x})$) are examined for this case.  The results are summarised in Fig.~\ref{figure_c} and Table \ref{Table_C}, and a conclusion similar as that in Sec.~\ref{Sec_case2D1} can be drawn.
\begin{table}[!ht]
  \centering
  \begin{tabular}{c|c|c|c|c}
  & Periodic& Cell manipulation & Cell transition & Both\\
  & structure& enabled only& enabled only & enabled\\
  \hline
   & & & &\\
  Compliance & 717.21 & 192.85 & 485.87 & 124.40\\
  \hline
  Volume& & & &\\
  fraction (\%) & 30.01 & 30.01 & 30.00 & 29.94\\
  \hline
  Optimised& & & &\\
  design & Fig.~\ref{figure_c}(a) & Fig.~\ref{figure_c}(b) & Fig.~\ref{figure_c}(c) & Fig.~\ref{figure_c}(d)\\
  \hline
  \end{tabular}
  \caption{Roles played by manipulation of a single cell and transition among generating cells coming from Eq.~\eqref{case2D2_chip}. Here cell manipulation involves cell rotation and distortion.\label{Table_C}}
\end{table}

\begin{figure}[!ht]
  \centering
  \includegraphics[width=.75\textwidth]{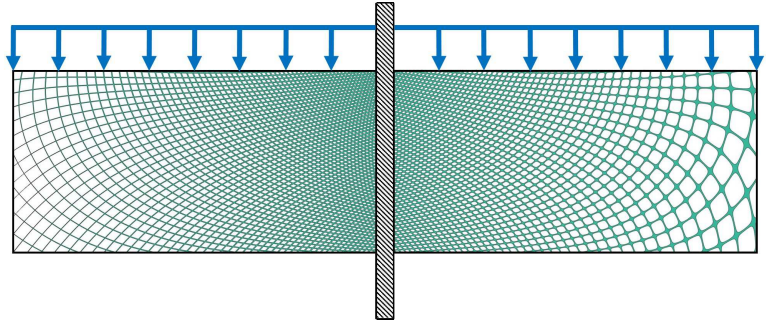}
  \caption{Left: the optimised IGM configuration from Fig.~\ref{uni_X_2D}. Right: the optimised IGM configuration from Fig.~\ref{uni_C_2D}.}
  \label{comparison_design_2D}
\end{figure}

Finally, we compare in Fig.~\ref{comparison_design_2D} the optimised designs based on the two sets of generating cells in Sec.~\ref{Sec_case2D1} and here, so as to summarise for some features of the present approach.
Several aspects of similarities are read. Firstly, microstructural conduits transiting the load to the fixed boundaries are seen in both optimised configurations. Secondly, the trends in volume change are the same. Thirdly, both optimised configurations output similar compliance values. But the design given by Fig.~\ref{uni_C_2D} tends to reveal how a trade-off is made during optimisation. It has been widely agreed that \cite{Allaire_book2002}, for compliance optimisation, laminate microstructures should deliver a near-optimal performance, provided that the laminate orientation properly align with the desired principal stresses. Compared to circular holes, configurations with a hyper-elliptic hole are morphologically closer to laminate structures. Fig.~\ref{uni_C_2D} indicates laminate microstructures are more desirable away from the fixed boundary and the loaded edge, while denser material distribution close to the fixed and loaded boundaries seems to be more effective for compliance optimisation.

\subsection{Three dimensional examples}
\begin{figure}[!ht]
  \centering
  \includegraphics[width=.65\textwidth]{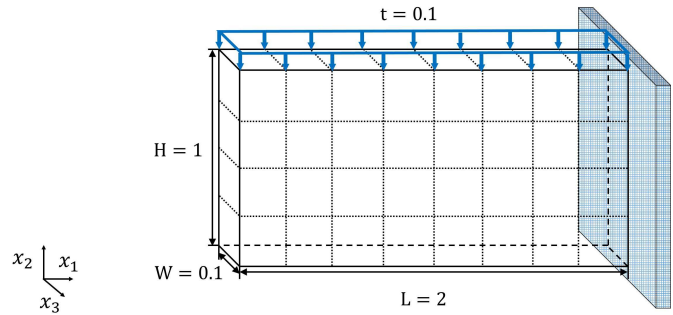}
  \caption{A three-dimensional short beam example.}
  \label{line_force_3D}
\end{figure}

In this subsection, topology optimisation results of three-dimensional cases are briefly presented, and further investigations are highly desired for its improvement. Here we consider a short beam of size $2\times1\times0.1$ with its right side fixed, as shown in Fig.~\ref{line_force_3D}.
On its upper surface, a uniformly distributed load is applied. The homogenised design domain is partitioned into $8\times4\times1$ identical cubic subdomains. For FE analysis, a mesh of $40\times20\times2$ is adopted.

\begin{figure}[!ht]
  \centering
  \includegraphics[width=.75\textwidth]{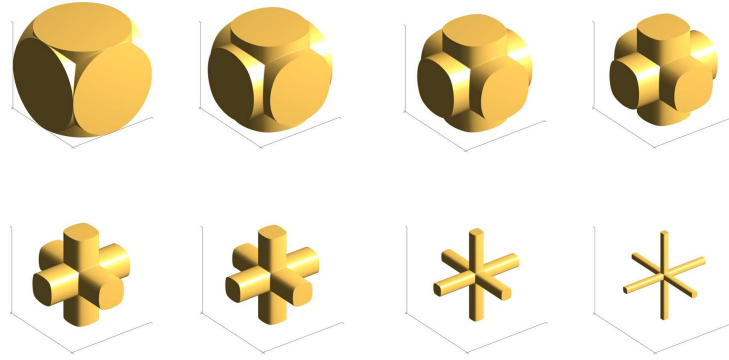}
  \caption{Samples from a continuous menu of generating cells: each cell is composed of three identical pillars of height 1, with each pillar crossing the other two. The cross-section of each pillar smoothly transits from a circle to a hyperellipse, as the volume fraction drops.}
  \label{initial_X_3D}
\end{figure}

The first example concerns a continuous menu of generating cells, whose representatives are shown in Fig.~\ref{initial_X_3D}.
Every cell consists of three identical pillars, with each pillar crossing the other two in their middles. The cross-sectional shape continuously changes from a unit circle to an almost diminishing hyper-ellipse, as the volume fraction drops. Mathematically, the solid parts of these generating cells can be described using a parameter $\zeta\in[0,1]$:
\begin{equation} \label{case3D1_cells}
\Upsilon_{\text{p}}^{\text{s}}(\zeta) = \bigcup_{\substack{i,j = 1 \\ i\neq j}}^{3} \left\{\bar{\mathbf{Y}} \,|\, (1-\zeta) \left(\bar{Y}_i^6 + \bar{Y}_j^6\right) + \zeta \left(\bar{Y}_i^2 + \bar{Y}_j^2\right) \le \frac{\zeta}{4},\, |\bar{\mathbf{Y}}|\le1\right\}.
\end{equation}

\begin{figure}[!ht]
  \centering
  \subfigure[Initial design]{\includegraphics[width=.38\textwidth]{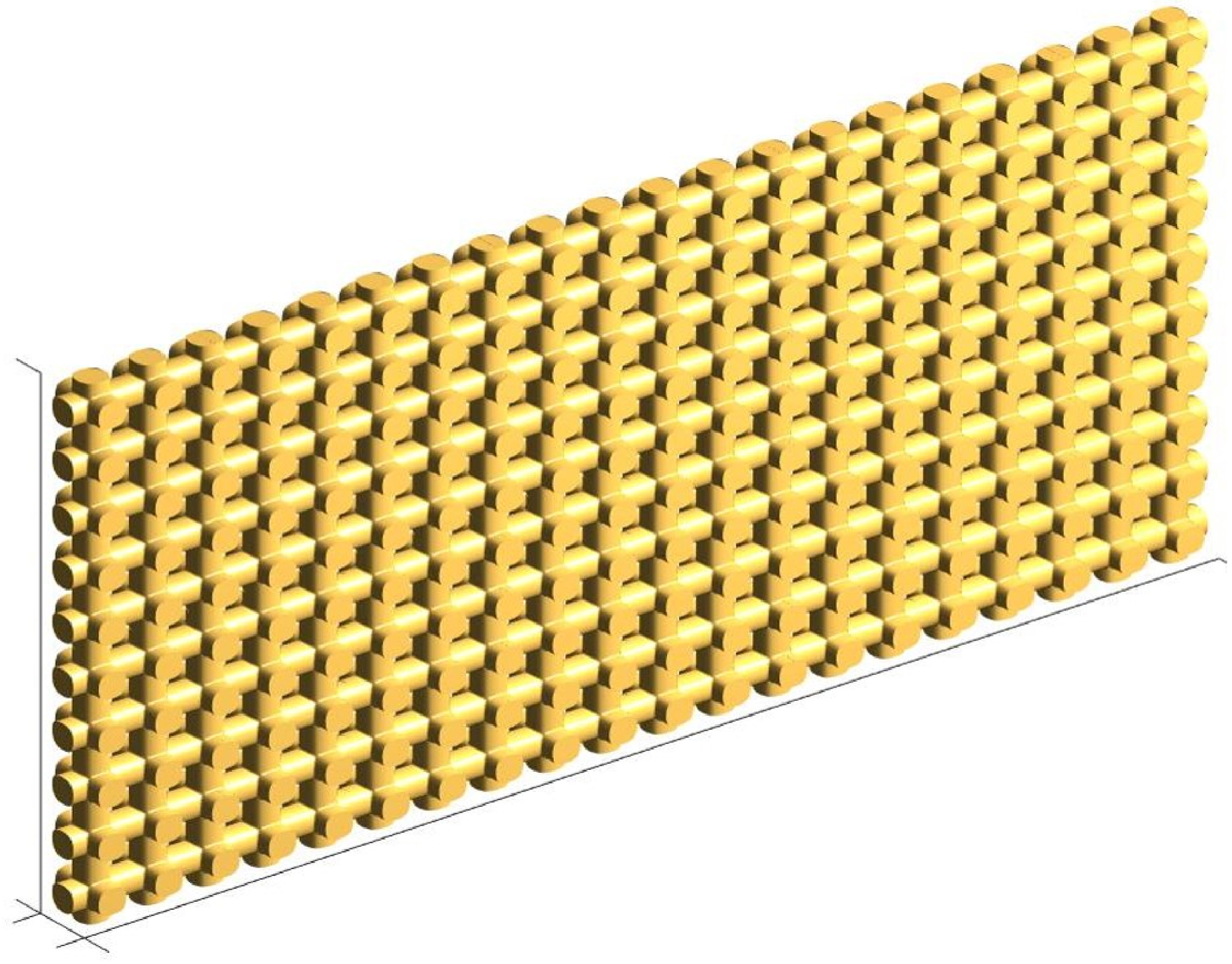}}
  \qquad
  \subfigure[Optimised design]{\includegraphics[width=.38\textwidth]{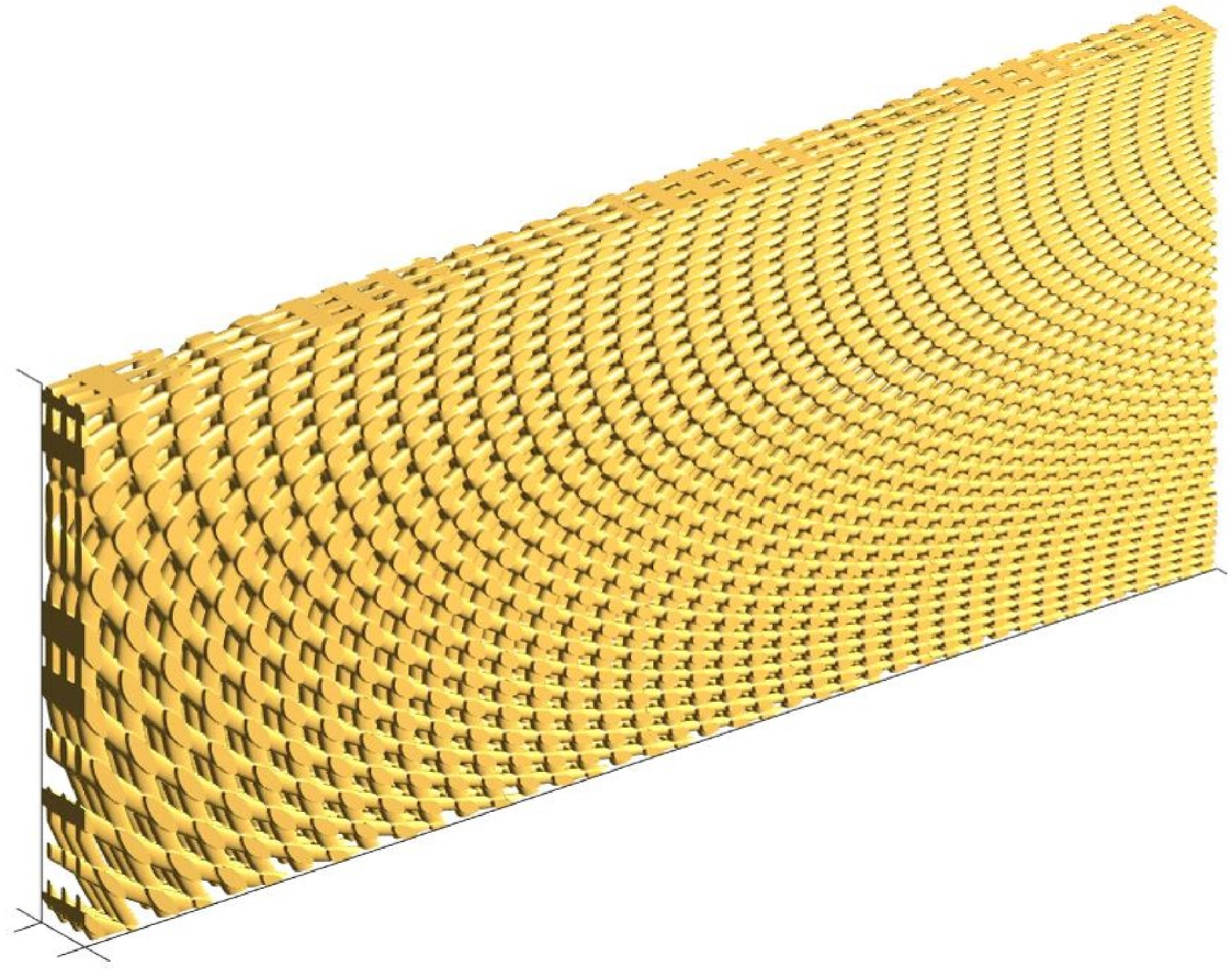}}
  \caption{The optimised design obtained based on a continuous menu of generating cells sampled in Fig.~\ref{case3D1_cells}.}
  \label{design_X_3D}
\end{figure}

Starting from an initially periodic structure shown in Fig.~\ref{design_X_3D}(a), the optimised IGM configuration is given in Fig.~\ref{design_X_3D}(b). In a side view, the three-dimensional optimised result bears a strong similarity with the corresponding two-dimensional configuration shown in Fig.~\ref{uni_C_2D}, that is, the microstructural cells arrange themselves for more effectively transiting the load to the fixed boundary.
Note that for better visualising the results, cell rotation is only permitted within the planes perpendicular to the $x_3$-direction of the design domain.

\begin{figure}[!ht]
  \centering
  \includegraphics[width=.75\textwidth]{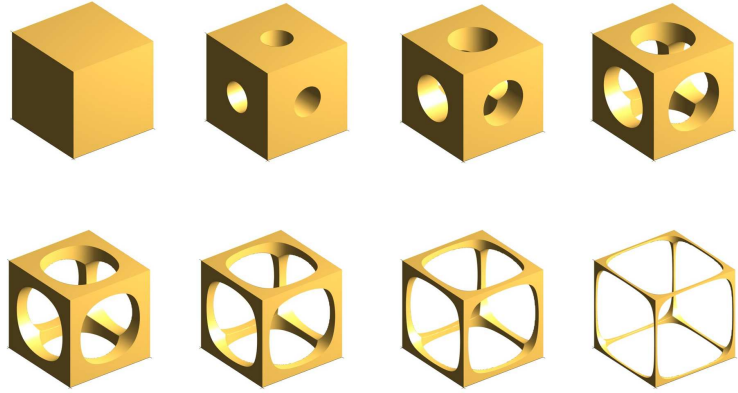}
  \caption{Samples from a continuous menu of generating cells: each cell is obtained by digging holes from a solid cube. The void domain consists of three cylindrical regions, crossing one and another. The cross-section of each cylinder smoothly transits from an almost diminishing circle to a hyperellipse of size 1, as the volume fraction drops.}
  \label{initial_O_3D}
\end{figure}

Another set of generating cells are also concerned with some of their representatives shown in Fig.~\ref{initial_O_3D}.
This time, every generating cell is obtained by digging holes from a solid cube. The void region can be envisaged as the union of three cylinders, one crossing the other two. The cross-section of each cylinder smoothly transits from an almost diminishing circle to a hyperellipse of size 1, as the volume fraction drops. Mathematically, the solid regions within the generating cells can be described by
\begin{equation} \label{case3D2_cells}
\Upsilon_{\text{p}}^{\text{s}}(\zeta) = \bigcup_{\substack{i,j = 1 \\ i\neq j}}^{3} \left\{\bar{\mathbf{Y}} \,|\, (1-\zeta) \left(\bar{Y}_i^2 + \bar{Y}_j^2\right) + \zeta \left(\bar{Y}_i^6 + \bar{Y}_j^6\right) \le \frac{\zeta}{64},\, |\bar{\mathbf{Y}}|\le1\right\}
\end{equation}
with $\zeta\in[0,1]$.

\begin{figure}[!ht]
  \centering
  \subfigure[Initial design]{\includegraphics[width=.38\textwidth]{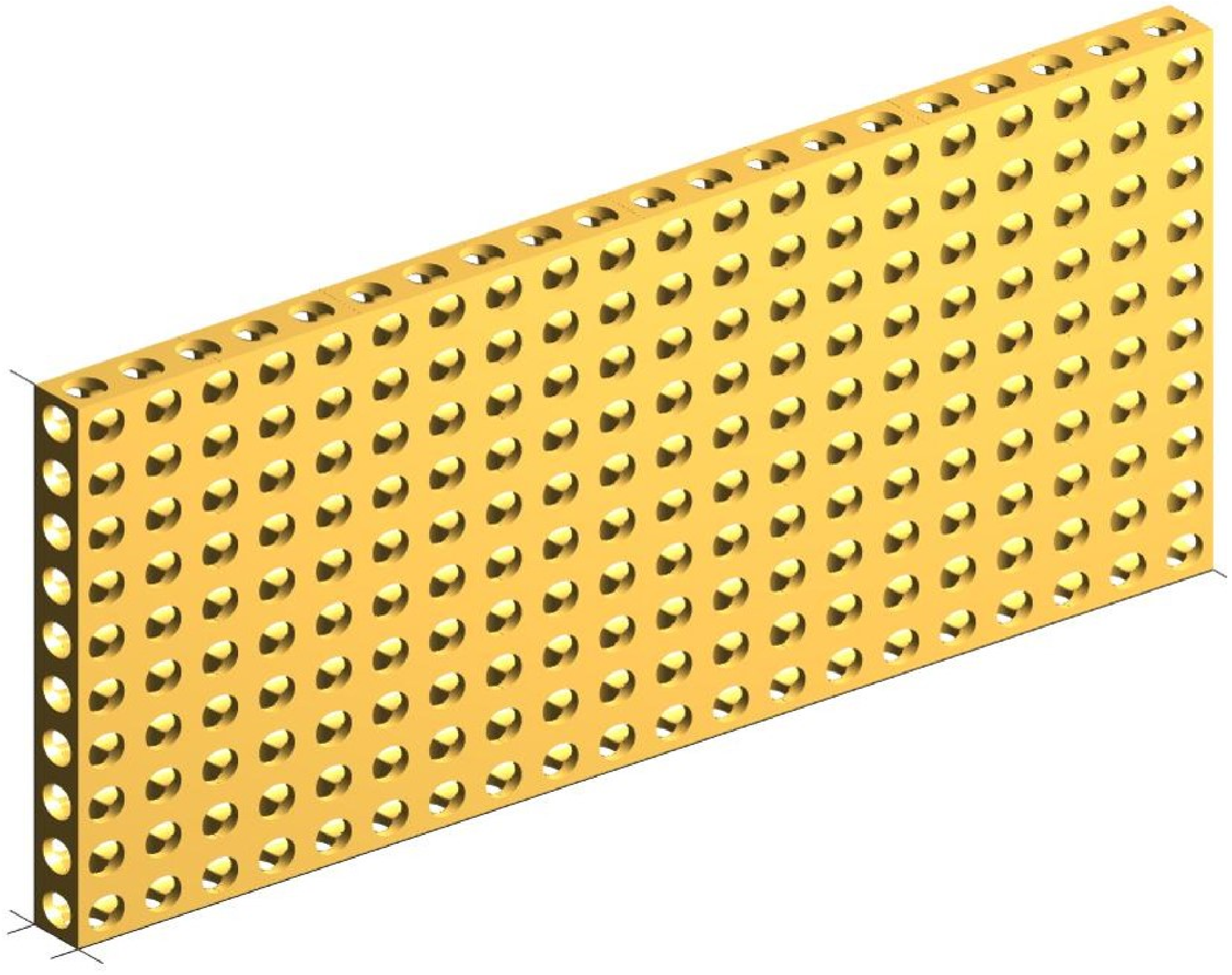}}
  \qquad
  \subfigure[Optimised design]{\includegraphics[width=.38\textwidth]{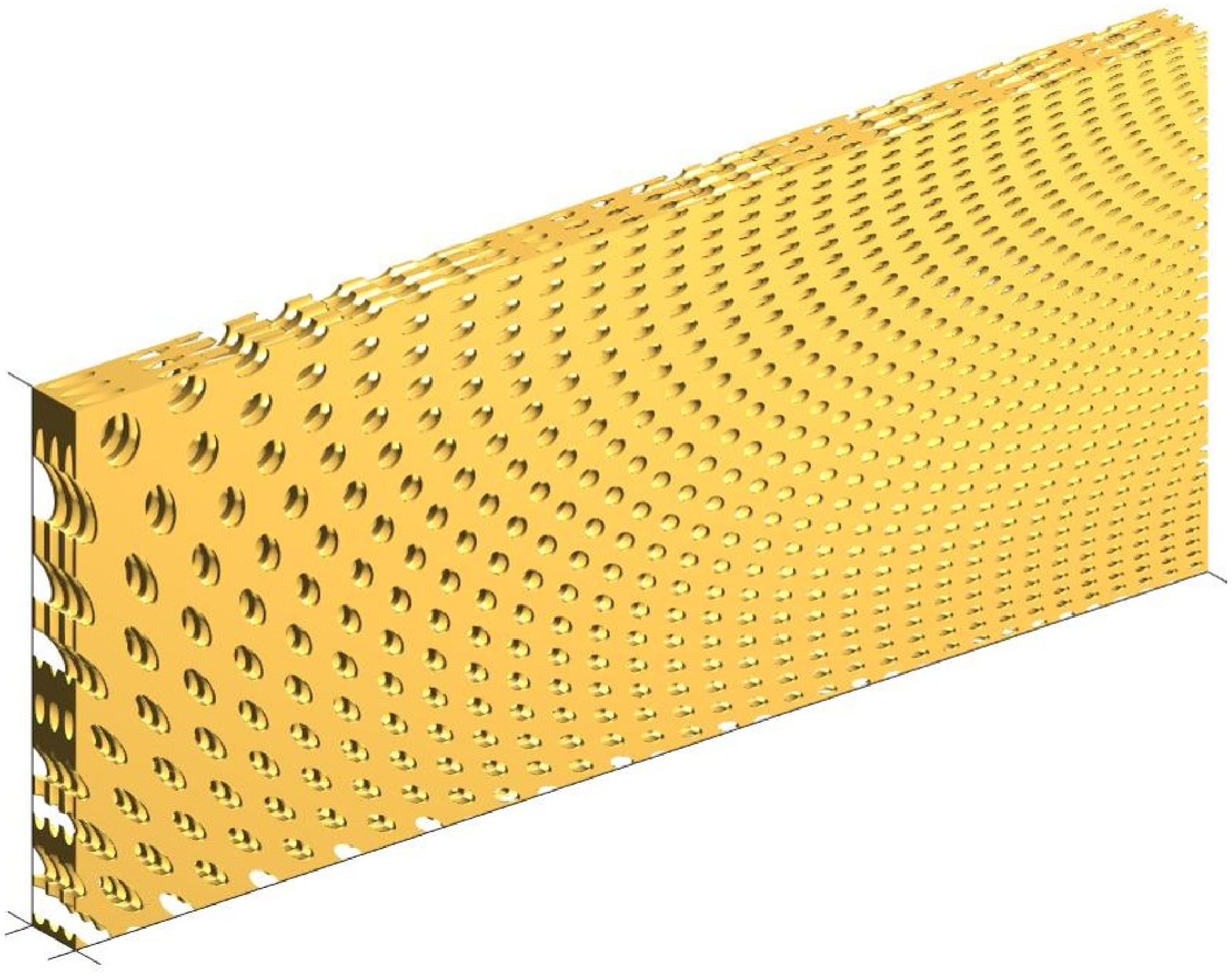}}
  \caption{The optimised design obtained based on a continuous menu of generating cells sampled in Fig.~\ref{initial_O_3D}.}
  \label{design_O_3D}
\end{figure}

Starting from an initially periodic configuration in Fig.~\ref{design_O_3D}(a), the optimised configuration is shown in Fig.~\ref{design_O_3D}(b), where the microstructural orientation obtained is roughly the same as that in Fig.~\ref{design_X_3D}(b).

\section{Conclusion and discussion\label{Sec_conclusion}}
The present article proposes an algorithm for generating infill graded microstructural configurations from a continuous menu of simple generating cells. With the present scheme, the IGM profile is controlled through the introduction of two functions: a level-set indicator function $\zeta(\mathbf{x})$ identifying which cell from the menu is of concern around a macroscopic point $\mathbf{x}$; and a mapping function $\mathbf{y}(\mathbf{x})$ indicating how the cell is rotated and distorted locally. With these two macroscopically continuous functions, IGM configurations can be generated based on TDF expressed by Eq.~\eqref{TDF_general}. Eq.~\eqref{TDF_general} effectively offers a unified expression where most existing methods for IGM generation find their traces, with detailed information summarised in Table~\ref{Table_comparison}. The present scheme also sees its generalised applications in two aspects: 1) IGM generation through interpolation among a set of desired seminal cell configurations; 2) IGM configurations decorated on a given manifold. The IGM mechanical behaviour can then be predicted using asymptotic analysis, and the accuracy of the homogenised formulation is further demonstrated through comparisons with the corresponding fine-scale simulation results. The sensitivity analysis is also studied, and a (homogenised) topology optimisation formulation for IGM design is also proposed. Numerical examples for two-dimensional cases are systematically analysed, and three-dimensional results are also briefly shown.

The present approach brings about considerable expansions in IGM describability under a topology optimisation framework underpinned by asymptotic analysis, which is termed as an AHTO plus framework here. Compared to its original version \cite{ZhuYC_JMPS2019}, the IGM configurations obtained from the present generation scheme inherit all the attractive IGM features, such as smooth connection and simple form in TDF expression. Moreover, IGM configurations with other practically useful features, such as smooth change of volume fraction in space, are also enabled by the present algorithm. As discussed in the beginning of this article, the original AHTO plus method \cite{ZhuYC_JMPS2019} needs essential improvements mainly over three issues: 1) to substantially raise the computational efficiency when implementing the homogenised formulation for stress analysis; 2) to overcome the limitation where all constituting cells are restricted to be self-similar to each other; 3) to find more appropriate basis functions governing the IGM generation through Eq.~\eqref{TDF_general}. The present article aims for addressing the second issue mentioned above, while the first issue is partly resolved by \cite{XueDC_arxiv2019} and the third issue still needs further investigations. This also points to one promising direction for future studies.

\section*{Acknowledgement}
The financial supports from National Key Research and Development Plan (2016YFB0201601) from the Ministry of Science and Technology of the People's Republic of China, the National Natural Science Foundation of China (11772076, 11732004, 11821202), Program for Changjiang Scholars, Innovative Research Team in University (PCSIRT) and 111 project (B14013) are gratefully acknowledged.


\bibliography{mybib}
\bibliographystyle{unsrt}




\end{document}